\documentclass[12pt]{article}
\usepackage[a4paper]{geometry}
\usepackage[utf8x]{inputenc}

\usepackage{graphicx, caption}
\usepackage{setspace}
\usepackage{bm}
\usepackage{amsmath}
\usepackage{siunitx}
\usepackage{caption}
\usepackage{subcaption}
\usepackage[backend=bibtex, style=nature]{biblatex}
\usepackage[font=scriptsize,skip=2pt]{caption}
\usepackage{authblk}
\usepackage{subfiles}
\usepackage{subcaption}
\usepackage{float}
\usepackage{overpic}
\usepackage{authblk}

\title{A novel understanding of the role of plasma-molecular kinetics on divertor power exhaust}

\author[1,3]{N.Osborne}
\author[2]{K.Verhaegh}
\author[1]{D.Moulton}
\author[1]{S.Mijin}
\author[4]{H.Reimerdes}
\author[1]{P.Ryan}
\author[1,5]{N.Lonigro}
\author[1]{R.Osawa}
\author[1,3]{K.Murray}
\author[2]{S.Kobussen}
\author[6]{Y.Damizia}
\author[4]{A.Perek}
\author[4]{C.Theiler}
\author[4]{R.Ducker}
\author[4]{D.Mykytchuk}

\author[a]{the Eurofusion Tokamak Exploitation Team}
\author[b]{the MAST-U Team}
\author[c]{the TCV Team}

\affil[1]{UKAEA, Culham Campus, Abingdon, OX14 3DB, UK}
\affil[2]{Eindhoven University of Technology, Eindhoven, the Netherlands}
\affil[3]{University of Liverpool, Liverpool, United Kingdom}
\affil[4]{Swiss Plasma Centre, École Polytechnique Fédérale de Lausanne, Lausanne, Switzerland}
\affil[5]{York Plasma Institute, University of York, York, United Kingdom of Great Britain and Northern Ireland}
\affil[6]{The College of William \& Mary, Williamsburg, VA, United States of America}
\affil[a]{See author list of E. Joffrin et al Nucl. Fusion 2024 10.1088/1741-4326/ad2be4}
\affil[b]{See author list of “J.R. Harrison et al 2019 Nucl. Fusion 59 112011}
\affil[c]{See author list of “H. Reimerdes et al 2022 Nucl. Fusion 62 042018}

\begin{document}
\date{}
\maketitle

\begin{abstract}
    During detachment, a buffer of neutral atoms and molecules builds up between the target and the ionising plasma. Collisions between the plasma and the molecules play an important role in the detachment process. Studies of plasma-molecular kinetics indicate that the gas temperature increases during deepening detachment for a wide range of conditions on the MAST-U and TCV tokamaks. This is related to an increased $\mathrm{D}_2$ lifetime during detachment, leading to more plasma-molecule collisions that raise the molecular temperature. Such collisions subsequently result in power and momentum losses from the divertor plasma during detachment. Using a simplified inference, these losses are estimated using the rotational temperature, neutral pressure and ionisation front position. Significant power losses (about $10\%$ of $P_{SOL}$) and dominant momentum losses (majority of the upstream pressure) from plasma-molecule collisions are inferred experimentally in long-legged, strongly baffled, detached divertors (MAST-U Super-X divertor).  These findings are consistent with SOLPS-ITER simulations (about $15\%$ of $P_{SOL}$, and $65\%$ of upstream pressure).  Simulations also suggest that $40–50\%$ of the power crossing the ionisation front is lost downstream.  The vibrational distribution obtained is compared with a collisional-radiative model setup using the same rate data as SOLPS-ITER, indicating some qualitative agreements and disagreements, potentially highlighting model gaps with regard to the default rates used.
    
    These interpretations highlight the importance of plasma-molecular collisions, leading to power and momentum losses during detachment.
\end{abstract}

\clearpage
\section{Introduction}\label{sec2_alt}

Plasma detachment will be a feature of future reactors to ensure that both divertor target electron temperature and heat flux are reduced by the orders of magnitude necessary \cite{Verhaegh2018, Stangeby2000}.  During detachment, plasma-neutral interactions (collisions and reactions) are key in the required removal of power, momentum and ions from the plasma \cite{Verhaegh2019, Verhaegh2021, Lipschultz1998, Stangeby2018, Krasheninnikov2017}.  Plasma-\textit{molecule} interactions, specifically, have now been shown to play a crucial role in detachment in several experimental devices \cite{Verhaegh2021b, Verhaegh2021a, Verhaegh2021, Verhaegh2023b}.\\

\begin{figure}[h!]
    \centering
    \includegraphics[width=0.5\linewidth]{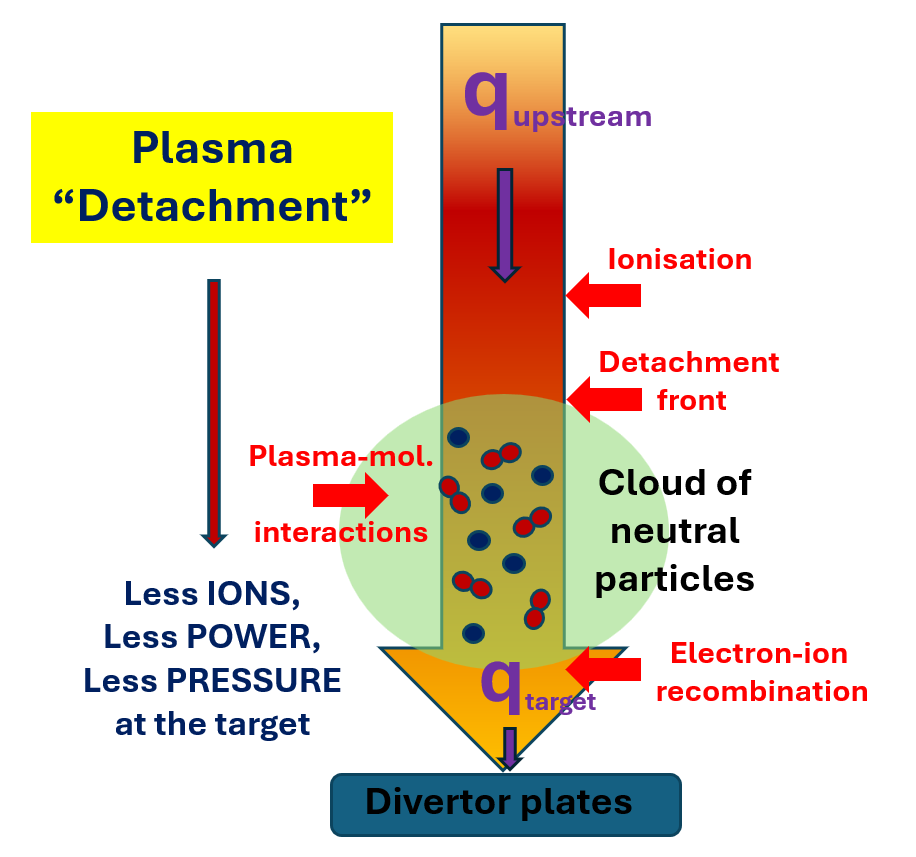}
    \caption{Simplified representation of the scrape-off-layer from mid-plane to divertor target in a deeply detached plasma.}
    \label{fig:detachment_cartoon}
\end{figure}

Figure \ref{fig:detachment_cartoon}, which shows an unfurled scrape-off-layer from mid-plane to divertor target, indicates the different processes present during detachment.  From upstream to downstream along the divertor leg, we pass from the ionisation region, through the neutral cloud where plasma-atom and plasma-molecule interactions occur, to the electron-ion recombination region closest to the target \cite{Verhaegh2023}.

The upstream edge of the neutral cloud, just below (downstream of) the ionisation region, corresponds with a peak in Fulcher emission from the molecules, as is well established by measurements and emission modelling in both MAST-U and TCV \cite{Verhaegh2023, Verhaegh2021b}.  Fulcher emission is a visible band of spectral lines between $\sim 600$ nm and $\sim 640$ nm emitted by electronically excited hydrogen molecules transitioning from the upper Fulcher state ($d^3\Pi^-_u$) to the lower Fulcher state ($a^3\Sigma^+_g$) and is very useful as a diagnostic tool.  It is at this edge where there starts to be a significant number of molecules but still enough hot electrons to electronically excite the molecules.  In this paper, we refer to this region of peak Fulcher emission as the ``detachment front" in keeping with other works where the same region is used to diagnose \cite{Verhaegh2023b,Wijkamp2023,Osborne2023} and even control \cite{BobK} detachment on MAST-U.\\

In detached conditions ($T_e <3-5$ eV), ion-molecule collisions have been found to be a dominant mechanism for power and momentum losses in MAST-U SOLPS-ITER simulations \cite{Omkar}.  Meanwhile, in the same detached conditions, reactions in the form of molecular chain chemistry processes are dominated by ion sinks (Molecular Activated Recombination or MAR) and neutral sources (Molecular Activated Dissociation or MAD) \cite{Verhaegh2021b, Verhaegh2024}.  This highlights the importance of plasma-molecule interactions in the detachment process.\\

The same $\mathrm{D}_2$ Fulcher emission which indicates the detachment front, can be monitored \textit{in high spectral resolution}, allowing for careful study of the molecules' rotational and vibrational, or ``rovibrational", distribution.  This measurement provides information on both collisions between the plasma and the molecules via the rotational distribution (rotational temperature); and information on the likelihood of the plasma to undergo molecular chain chemistry processes, which favour highly vibrationally excited molecules, via the vibrational distribution.\\

Diagnosing the rovibrational distribution of $\mathrm{D}_2$ molecules is, therefore, potentially very useful in providing insight into the plasma-molecule interactions which are so fundamental in detachment.\\

Specifically, we have found that increasing rotational, or gas, temperature is a signature of energy transfer from the plasma to the molecules through collisions, the same process that drives momentum losses during detachment.  Meanwhile, the vibrational distribution directly drives effective molecular charge exchange rates and affects molecular ionsiation rates \cite{Verhaegh2023a,Chandra2023,Holm2022}.  These are the first stages in the MAR and MAD processes described above and, hence, the vibrational distribution plays a role in these processes.\\

\subsection{Historical work}
Investigations on tokamaks (JET \cite{Sergienko2013MolecularJET}, DIII-D \cite{Hollmann2006}, ASDEX-Upgrade \cite{Fantz2001} and JT60-U \cite{Kubo2005}), and other devices such as LHD \cite{Fujii2023Plasma-parameterDivertor} and Magnum-PSI \cite{Akkermans2020}, have inferred rotational temperatures for the ground-state of $\mathrm{D}_2$ between approximately 1000 K and 10000 K.  Measurements on JET \cite{Sergienko2013MolecularJET}, DIII-D \cite{Hollmann2006} and LHD \cite{Fujii2023Plasma-parameterDivertor}, suggest a correlation of rotational temperature with electron density, and \citeauthor{Hollmann2006} \cite{Hollmann2006} shows that, at the same electron density on DIII-D, TEXTOR, and PISCES (linear device), similar rotational temperatures are measured.  Recent work on two metal-walled devices (QUEST and LTX-$\beta$) and carbon-walled DIII-D by \citeauthor{Yoneda2023SpectroscopicTokamaks} \cite{Yoneda2023SpectroscopicTokamaks} suggests a high sensitivity of rotational temperature to electron density and attempts to predict the rotational temperature increases by use of a collisional-radiative model with 858 differential equations.  The same study indicated a low sensitivity to electron temperature, at least in plasma regimes above 10 eV.\\

Previous experiments \cite{Fantz1998SpectroscopicMolecules, Briefi2020APlasmas, Hollmann2006, Brezinsek2002MolecularTEXTOR-94, Brezinsek2005} inferred a vibrational temperature of the molecular ground-state by using Franck-Condon coefficients to map an assumed vibrational temperature in the ground state to the electronically-excited Fulcher level. Results on JET-ILW \cite{Sergienko2013MolecularJET}, however, feature an overpopulated $\nu=0$ band in the excited Fulcher level, suggesting a strongly non-Boltzmann ground state with highly overpopulated $\nu=1,2,3$.  TEXTOR results (carbon surface) \cite{Brezinsek2005} found an overpopulation of the $\nu=3$ band and thus an imperfect fit with a ground state Boltzmann distribution.\\

Our previous work \cite{Osborne2023}, studied the MAST-U \cite{MU-Device} divertor in Ohmic density ramp discharges and found rotational temperatures consistent with these previous findings.  The rotational temperature was found to increase in deep detachment in the strongly baffled MAST-U divertor, hinting at a connection to longer lifetimes of the molecules before dissociation or ionisation.  Our MAST-U results also demonstrated a non-Boltzmann distribution for the ground-state vibrational distribution of the molecules, and revealed a curious boost to the $\nu=2,3$ populations during detachment \cite{Osborne2023}.\\

\subsection{This paper}
In this work high spectral resolution $\mathrm{D}_2$ $\&$ $\mathrm{H}_2$ Fulcher band emission is obtained in long-legged divertors of both MAST-U and TCV \cite{TCV-Device} to extract the rotational and vibrational distribution of the molecules.\\

We investigate how information from the \textbf{rotational distribution} of $\mathrm{D}_2$ $\&$ $\mathrm{H}_2$ can provide more insight into the kinetics of the molecules in a long-legged divertor and what those kinetics imply for the divertor state in terms of energy and momentum transfer from the plasma to the molecules.  Meanwhile, we investigate the \textbf{vibrational distribution} of $\mathrm{D}_2$ $\&$ $\mathrm{H}_2$, noting its bearing on plasma chemistry, and compare it to distributions predicted by collisional-radiative models and inherently used in high-fidelity simulation codes.\\

Going beyond our previous ohmic MAST-U work \cite{Osborne2023}, this study uses the unique features of both the MAST-U and TCV tokamaks to obtain improved insights into what impacts plasma-molecular kinetics. The following features are varied:

\begin{itemize}
        \item Divertor configuration, e.g. variations in poloidal leg length and total flux expansion (MAST-U), to investigate the impact of divertor leg length on plasma-molecular kinetics.
        \item Baffled and unbaffled conditions (TCV), to investigate the impact of neutral closure and trapping in the divertor on plasma-molecular kinetics.
        \item Fuelling location and fuelling strength (TCV and MAST-U), to investigate their impact on plasma-molecular kinetics.
        \item Fuelling species (isotopic variation) ($\mathbf{H}_2$ in TCV), to investigate the impact of different isotopes on plasma-molecular kinetics.
        \item External heating levels (MAST-U with/without NBI heating) to investigate the impact of external heating on plasma-molecular kinetics in the divertor.
        \item L- and H-mode (MAST-U), to investigate whether regular transients (ELMs in H-mode) have an impact on plasma-molecular kinetics in the inter-ELM/'baseline' state.
\end{itemize}

We find that the observed behaviour of the rotational and vibrational distribution of hydrogenic molecules, in response to the above variations, is similar on MAST-U and TCV. For all conditions, the rotational temperature rises during and near detachment. For both TCV and MAST-U, this increase in rotational temperature is correlated with an enhancement of the $\nu=2,3$ population in the upper Fulcher state.\\

This work provides novel insights into the importance of plasma-molecule collisions on detachment.  A reduced model explains the observed rotational temperature increase, consistent with SOLPS-ITER simulations: the energy transfer from the plasma to the molecules is enhanced in detached conditions as the lifetime of the molecules is increased. Collisions between the plasma and the molecules result in significant power loss (up to $>10\%$ of $P_{SOL}$, the power entering the Scrape-Off-Layer); and losses of up to $>40\%$ downstream of the ionisation front); and very significant momentum sinks.\\

The measured vibrational distribution is yet to be accurately reproduced by collisional-radiative modelling which employ the same plasma-molecular rates used in exhaust simulations.  This mismatch increases uncertainty in the rates used, and motivates more experimental and modelling studies.
\section{Experimental Setup}\label{Experimental Setup}

Both MAST-U and TCV have clean intersections between the spectroscopic lines-of-sight and the divertor plasma.  Furthermore, both devices support long divertor legs, enabling measurements of the rotational temperature along the divertor leg.  \\

\begin{figure}[ht]
  \centering
  \includegraphics[width=\linewidth]{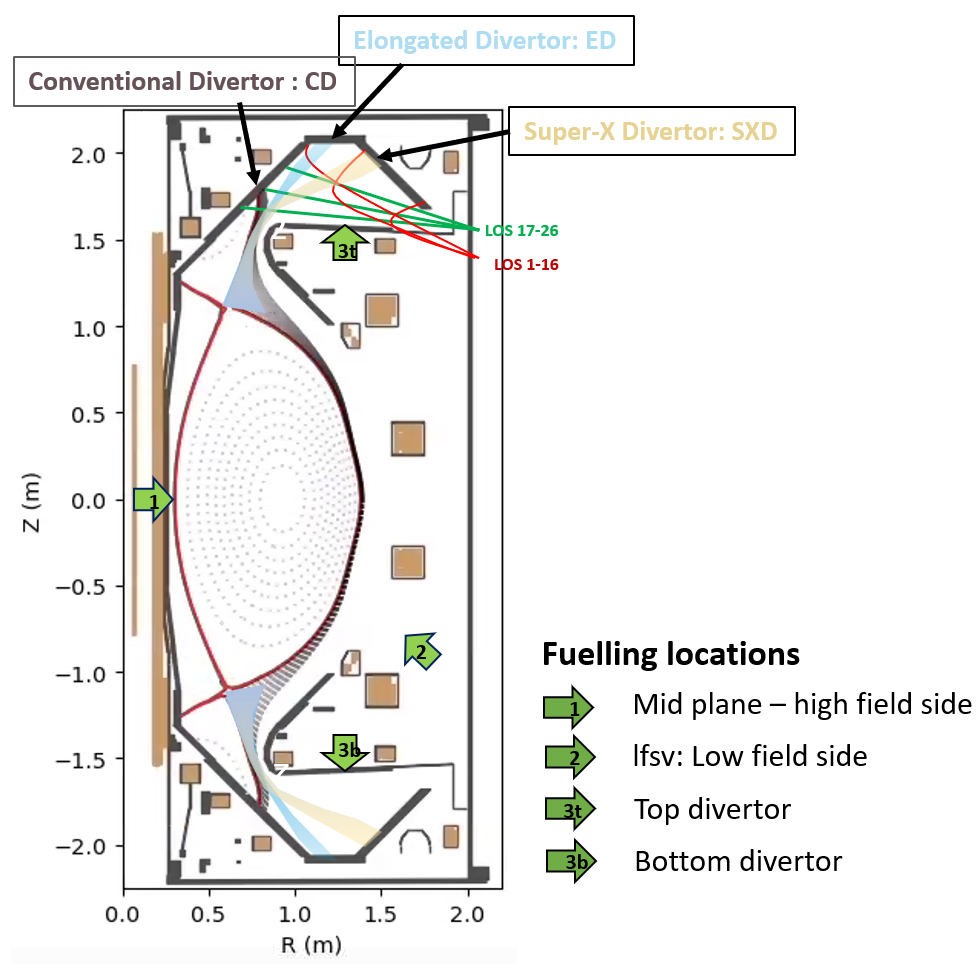}
  \caption{Schematic showing the divertor configurations, lines of sight and fuelling locations in the MAST-U upper divertor.}
  \label{MU_schematic}
\end{figure}

\begin{figure}[ht]
  \centering
  \includegraphics[width=0.8\linewidth]{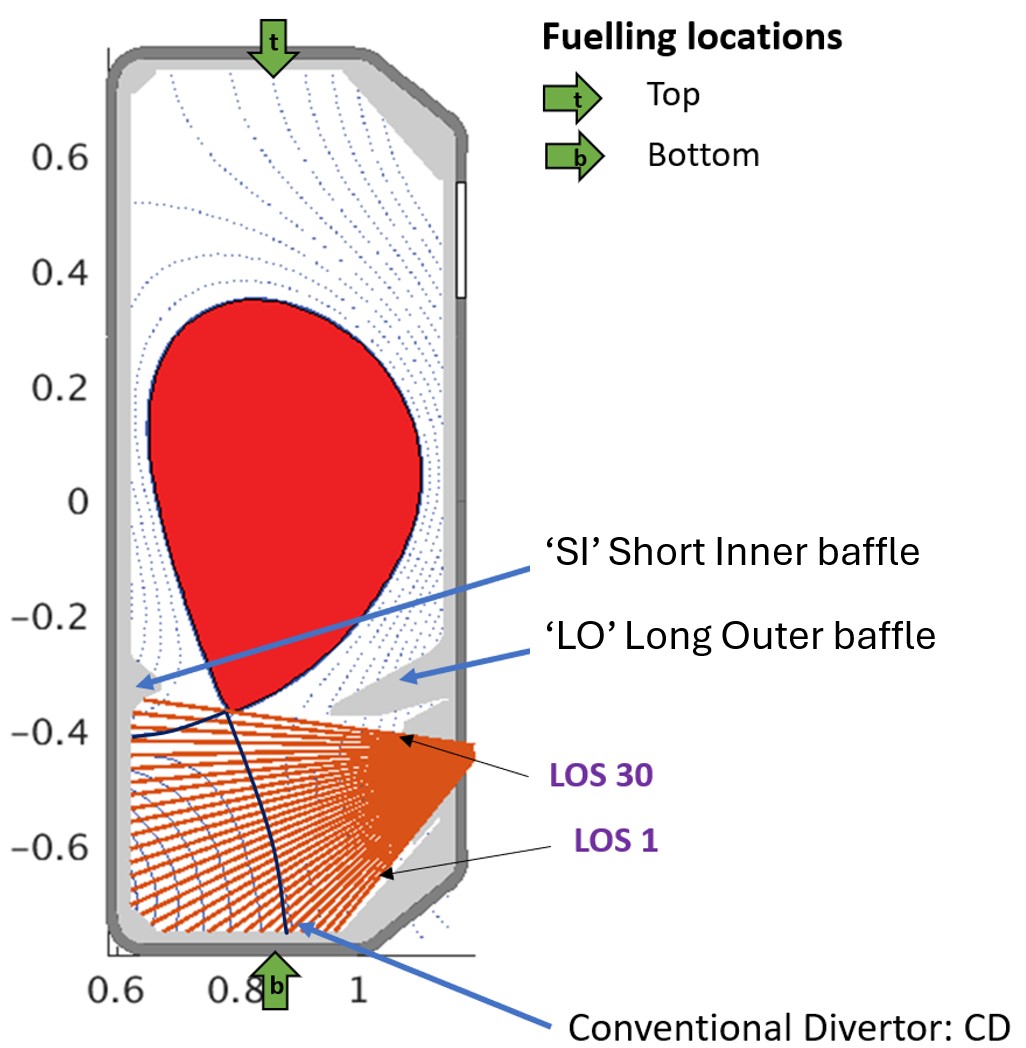}
  \caption{Schematic showing the divertor configuration, lines of sight and fuelling locations in the TCV divertor.}
  \label{TCV_schematic}
\end{figure}

Tables \ref{tab:MU discharges} and \ref{tab:TCV discharges} summarise the MAST-U and TCV discharges examined in this paper. These generally either contain density ramp experiments (wide range of core densities, expressed as a fraction of the core Greenwald density, $f_{GW}$,) or experiments with as constant a range of core density as possible with divertor fuelling (narrow $f_{GW}$ range).  The configurations and fuelling locations referred to are shown in figures \ref{MU_schematic} and \ref{TCV_schematic}.  The MAST-U discharges were beam-heated ($P_{NBI} > 1.5$ MW), while the TCV discharges were Ohmic.  All the discharges were unseeded.\\

\begin{table}[ht]
    \centering
    \begin{tabular}{|>{\centering\arraybackslash}p{2cm}|>{\centering\arraybackslash}p{1.5cm}|>{\centering\arraybackslash}p{3.5cm}|>{\centering\arraybackslash}p{1cm}|>{\centering\arraybackslash}p{1.5cm}|>{\centering\arraybackslash}p{2.5cm}|>
    {\centering\arraybackslash}p{1.5cm}|}
        \hline
        \textbf{Discharge} & \textbf{Config.} & \textbf{Fuelling} & \textbf{Mode} & \textbf{P\textsubscript{SOL}\textsuperscript{max}} & \textbf{f\textsubscript{GW}} & \textbf{Isotope} \\
        \hline
        49408 & CD & 2 & L & 1.3 MW & 0.25 - 0.50 & D\textsubscript{2} \\
        48008 & ED & 2 & L & 1.3 MW & 0.20 - 0.47 & D\textsubscript{2} \\
        48358 & SXD & 2 & L & 1.3 MW & 0.27 - 0.49 & D\textsubscript{2} \\
        49286 & SXD & 2 + puffs: 3b, 3t & L & 1.1 MW & 0.25 - 0.32 & D\textsubscript{2} \\
        49289 & SXD & 2, 3b, 3t & L & 1.2 MW & 0.25 - 0.37 & D\textsubscript{2}\\
        49324 & SXD & 1, 3b, 3t & H & 2.8 MW & 0.49 - 0.596 & D\textsubscript{2}\\
        \hline
    \end{tabular}
    \vspace{0.5cm} 
    \caption{MAST-U discharges examined in this paper.  The configurations and fuelling locations referenced can be seen in figure \ref{MU_schematic}.}
    \label{tab:MU discharges}
\end{table}

\begin{table}[ht]
    \centering
    \begin{tabular}{|>{\centering\arraybackslash}p{2.3cm}|>{\centering\arraybackslash}p{4cm}|>{\centering\arraybackslash}p{1.4cm}|>{\centering\arraybackslash}p{1cm}|>{\centering\arraybackslash}p{1.7cm}|>{\centering\arraybackslash}p{1.9cm}|>
    {\centering\arraybackslash}p{1.3cm}|}
        \hline
        \textbf{Discharge} & \textbf{Config.} & \textbf{Fuelling} & \textbf{Mode} & \textbf{P\textsubscript{SOL}\textsuperscript{max}} & \textbf{f\textsubscript{GW}} & \textbf{Isotope} \\
        \hline
        78024/78042 & CD + ``SILO" baffling  & b & L & 0.35 MW & 0.18 - 0.54 & D\textsubscript{2} \\
        76113/76114 & CD + \,\,\,\,\,\,\,\,\,\,\,\,no baffling & b & L & 0.32 MW & 0.18 - 0.57 & D\textsubscript{2} \\
        78045 & CD + ``SILO" baffling & t & L & 0.4 MW & 0.19 - 0.54 & D\textsubscript{2} \\
        78047 & CD + ``SILO" baffling & b & L & 0.36 MW & 0.2 - 0.44 & D\textsubscript{2} \\
        78440/78441 & CD + ``SILO" baffling & b & L & 0.4 MW & 0.2 - 0.49 & H\textsubscript{2}\\
        \hline
    \end{tabular}
    \vspace{0.5cm} 
    \caption{TCV discharges examined in this paper.  When two discharges are shown, they are repeat shots to enable full spectrometer coverage of the Fulcher band.  The fuelling locations can be referenced in figure \ref{TCV_schematic} as can the baffling description.}
    \label{tab:TCV discharges}
\end{table}

Figure \ref{MU_schematic} shows a schematic of the MAST-U tokamak device and indicates the three divertor configurations: Conventional (CD), Elongated (ED) and Super-X (SXD); the spectrosopic lines of sight in the upper divertor; and the fuelling locations relevant to the discharges examined in this paper.\\

Figure \ref{TCV_schematic} shows a schematic of the TCV tokamak device in the conventional divertor (CD) configuration indicating the spectroscopic lines of sight and the fuelling locations relevant to the discharges examined in this paper.\\

In order to observe the evolution of the molecular rovibrational distribution with detachment, some meaningful measure of detachment is required.  For this purpose, we adopt the peak ion target flux near the strike point (ions/$m^2$/$s$) since it is observed that the rollover of the peak ion flux corresponds well to the separation of the Fulcher front (and therefore ionisation front) from the target \cite{Verhaegh2023}.  Spectroscopic lines of sight near the strike point are selected for extracting the rovibrational information.\\

\subsection{Spectroscopic analysis and extracting spectroscopic information}

The rovibrational distribution is obtained using high resolution Fulcher band analysis using the technique described in detail in \cite{Osborne2023}. The rotational temperatures of $\mathrm{D_2}$ and $\mathrm{H_2}$ quoted in these results refer to the rotational temperature of the $\nu=0$ band of the \textit{ground} state.  Rotational temperatures are obtained from the distribution of rotational energy levels in the ``upper Fulcher state", or $d^3\Pi^-_u$, experimentally observed via spectroscopic lines produced by spontaneous electronic decay from $d^3\Pi^-_u$ to $a^3\Sigma^+_g$.  This Boltzmann distribution of rotational states is mapped to the ground state via a simple scale factor which relates the rotational temperature of the upper Fulcher state to the ground state. \\

Under conditions where the ground state is long-lived in comparison with characteristic collision time, then it is generally assumed that the rotational temperature of molecules in their electronic ground state can be used as a proxy for the kinetic, or `gas', temperature within 10-20\% \cite{Trot-Tgas-Astashkevich}.  We find that the rotational states of the first vibrational band in the upper Fulcher state, a mapping of the distribution in the ground state, consistently form a good Boltzmann distribution in measurements.  This suggests that collisionality is indeed high enough for equilibration between rotational and kinetic states.\\

\begin{figure}[ht]
  \centering
  \begin{subfigure}{\linewidth}
    \centering
    \includegraphics[width=1\linewidth]{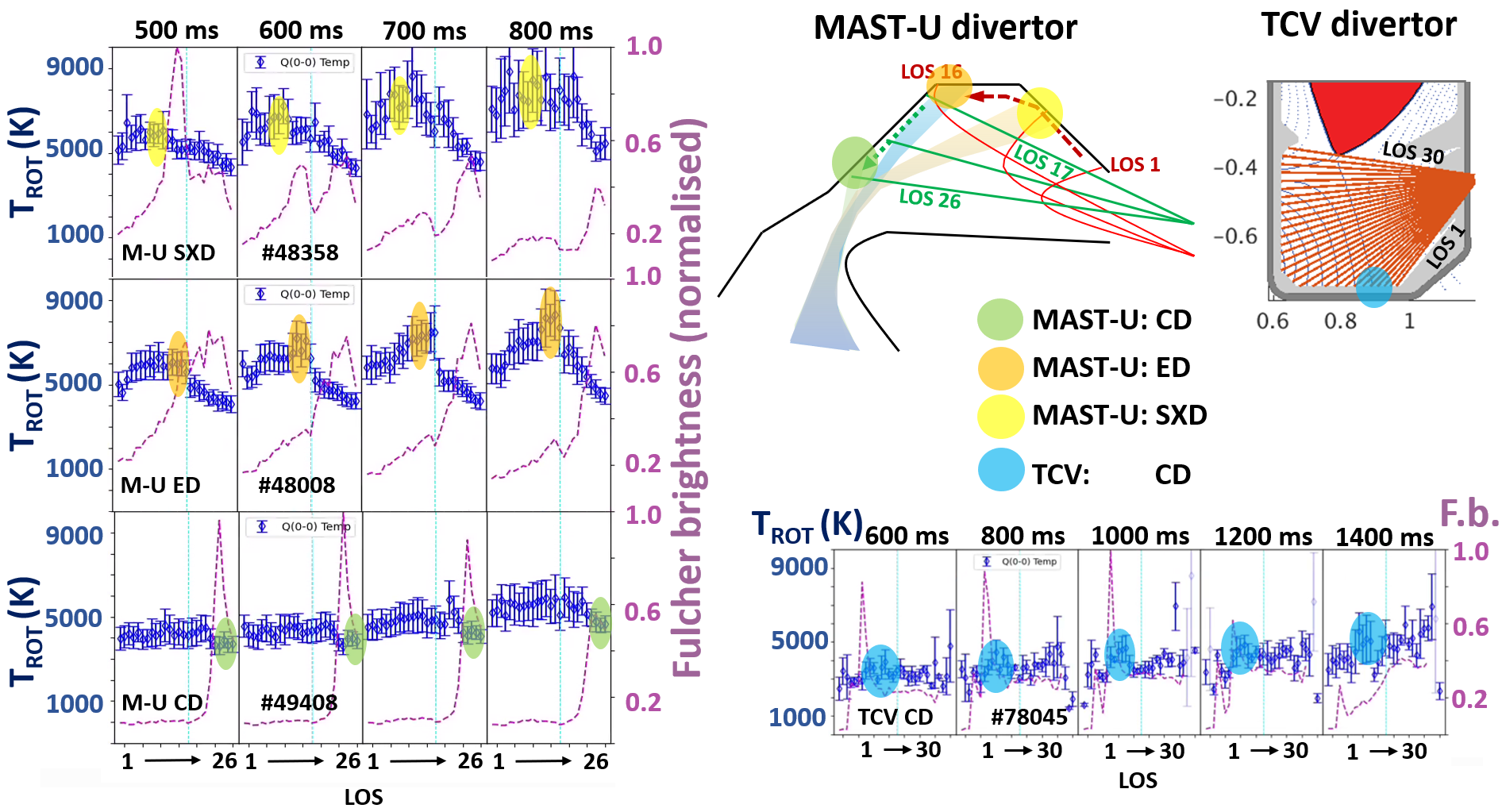}
    \begin{picture}(0,0)
      \put(-230,235){\textbf{(a)}}
      \put(1,235){\textbf{(b)}}
      \put(1,110){\textbf{(c)}}
    \end{picture}
  \end{subfigure}
  \caption{The areas of interest for each divertor configuration and imaging scans. The graphs in (a) show examples of the evolution of rotational temperature ($T_{ROT}$) across the divertor for an SXD, ED, and CD configuration in MAST-U.  (c) shows the evolution of $T_{ROT}$ for TCV (CD configuration).  (b) shows the lines of sight (LOS) in the divertors of MAST-U and TCV and the highlighted colour circles indicate the grouping of LOS near the strike point in all configurations.  These highlights correspond to the highlighted LOS in (a) and (c).}
  \label{sp_groups}
\end{figure}

Examples of spatial profiles of rotational temperature evolving during a discharge are shown in figure \ref{sp_groups}a for the MAST-U conventional, elongated and Super-X divertor configurations, and in figure \ref{sp_groups}c for the TCV conventional divertor configuration.  The spectroscopic lines of sight in the MAST-U and TCV divertors are also shown in figure \ref{sp_groups}b.  The highlighted coloured blocks in both sets of diagrams show the groups of lines of sight near the strike-point.  The rotational temperatures obtained at these locations are used to enable comparisons against multiple conditions/divertor shapes/devices.\\

\begin{figure}
  \begin{subfigure}{\linewidth}
    \centering
    \includegraphics[width=1\linewidth]{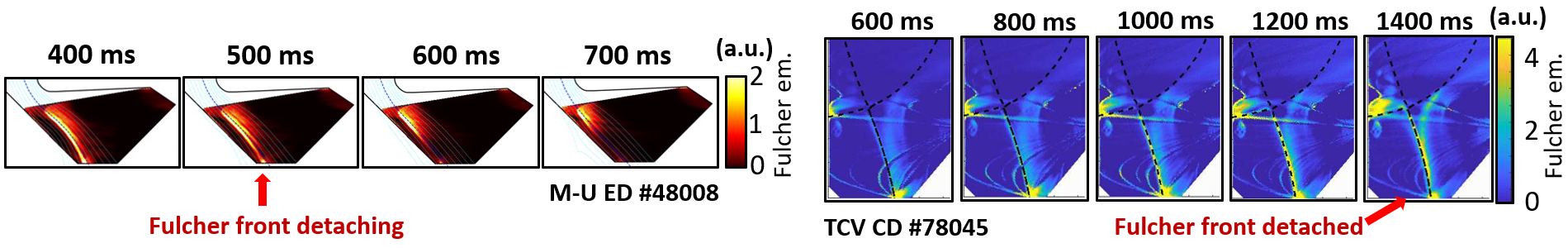}
    \begin{picture}(0,0)
      \put(-230,80){\textbf{(a)}} 
      \put(1,80){\textbf{(b)}}
    \end{picture}
  \end{subfigure}
    \caption{Fulcher emission imaging scans at various times during (a) the ED MAST-U discharge 48008 and (b) the TCV discharge 78045.  (Note in (b) the bright spot seen at the target after detachment is partially due to an inversion artifact and partially an impurity spectral line.)}
  \label{mwi_groups}
\end{figure}

Figure \ref{mwi_groups}a and \ref{mwi_groups}b show examples of imaging of the Fulcher emission in MAST-U and TCV corresponding to figures \ref{sp_groups}a (ED) and \ref{sp_groups}c respectively.  The Fulcher emission front (an indicator of the ionisation front) is seen to detach from the target at around 500 ms in the MAST-U case, and after 1200 ms in the TCV case.  These can be compared to peak particle flux (see figures \ref{SP_graphs}f and \ref{SP_graphs}c respectively to observe the correlation with peak ion flux rollover).\\

\section{Results}\label{Results}

\subsection{Rotational temperature evolution at the strike point}

Using variations in 1) divertor configuration (MAST-U); 2) neutral baffling (TCV); 3) fuelling location (TCV); 4) fuelling isotope (TCV) and 5) external heating levels (MAST-U) the rotational temperature near the strike point is investigated as function of core density. The core density is gradually increased using feed-back control to obtain density ramps. \\

\begin{figure}
    \centering
    \begin{minipage}[b]{0.47\textwidth}
        \centering
        \begin{overpic}[width=1.05\textwidth]{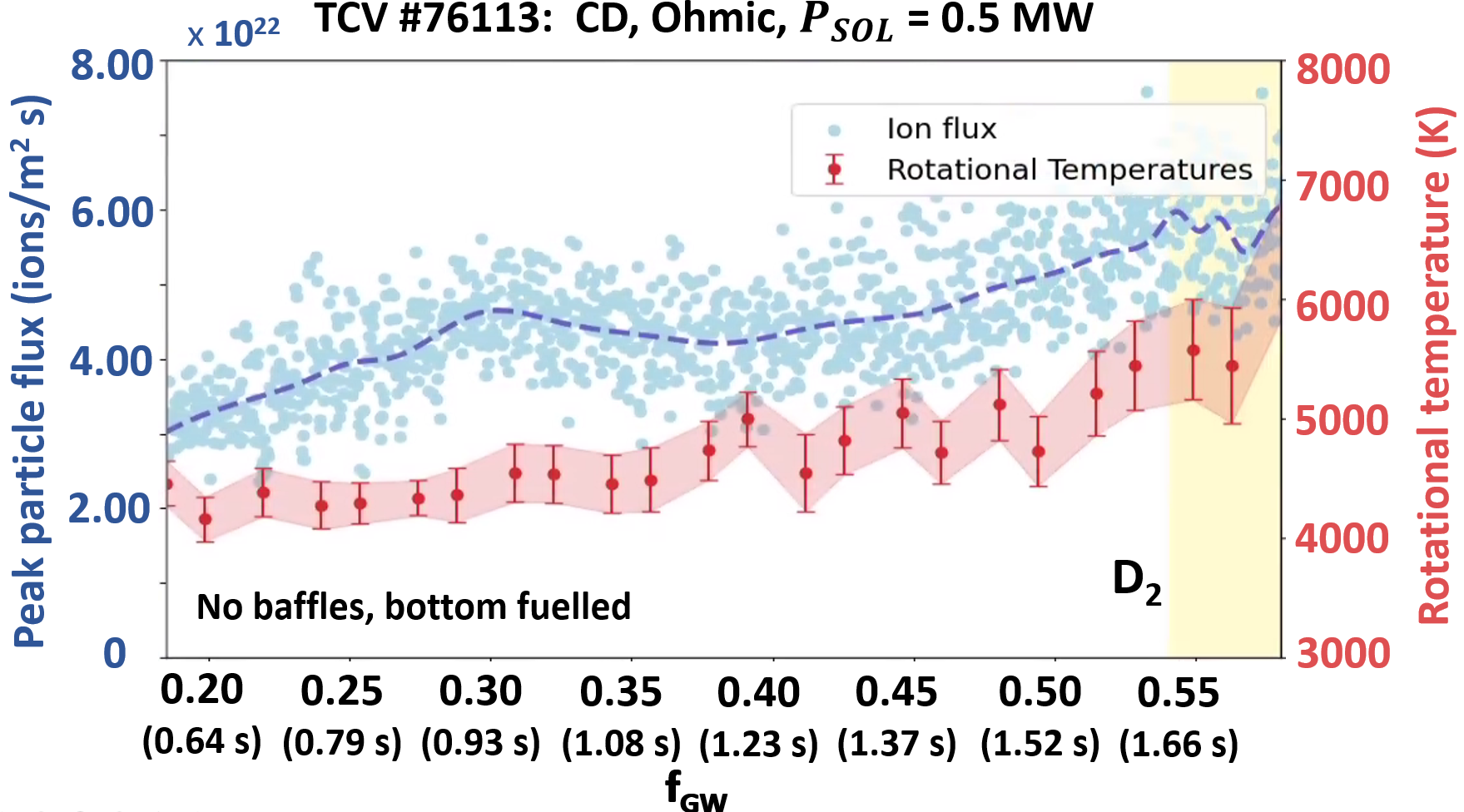}
            \put(15,47){\small \textbf{(a)}}
        \end{overpic}
    \end{minipage}
    \hspace{0.02\textwidth}
    \begin{minipage}[b]{0.47\textwidth}
        \centering
        \begin{overpic}[width=1.05\textwidth]{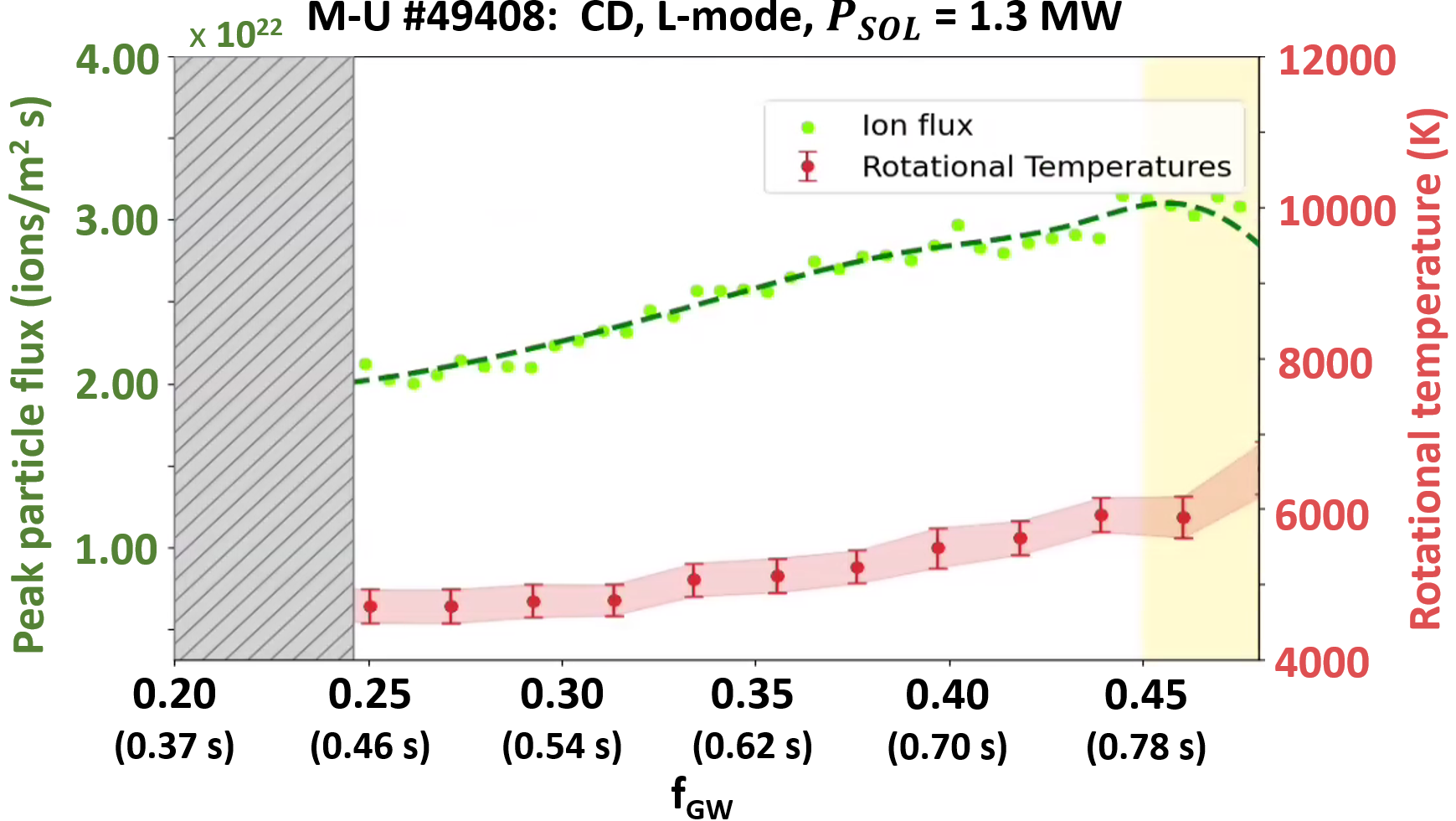}
            \put(15,47){\small \textbf{(e)}}
        \end{overpic}
    \end{minipage}
    \vspace{1em}

    \begin{minipage}[b]{0.47\textwidth}
        \centering
        \begin{overpic}[width=1.05\textwidth]{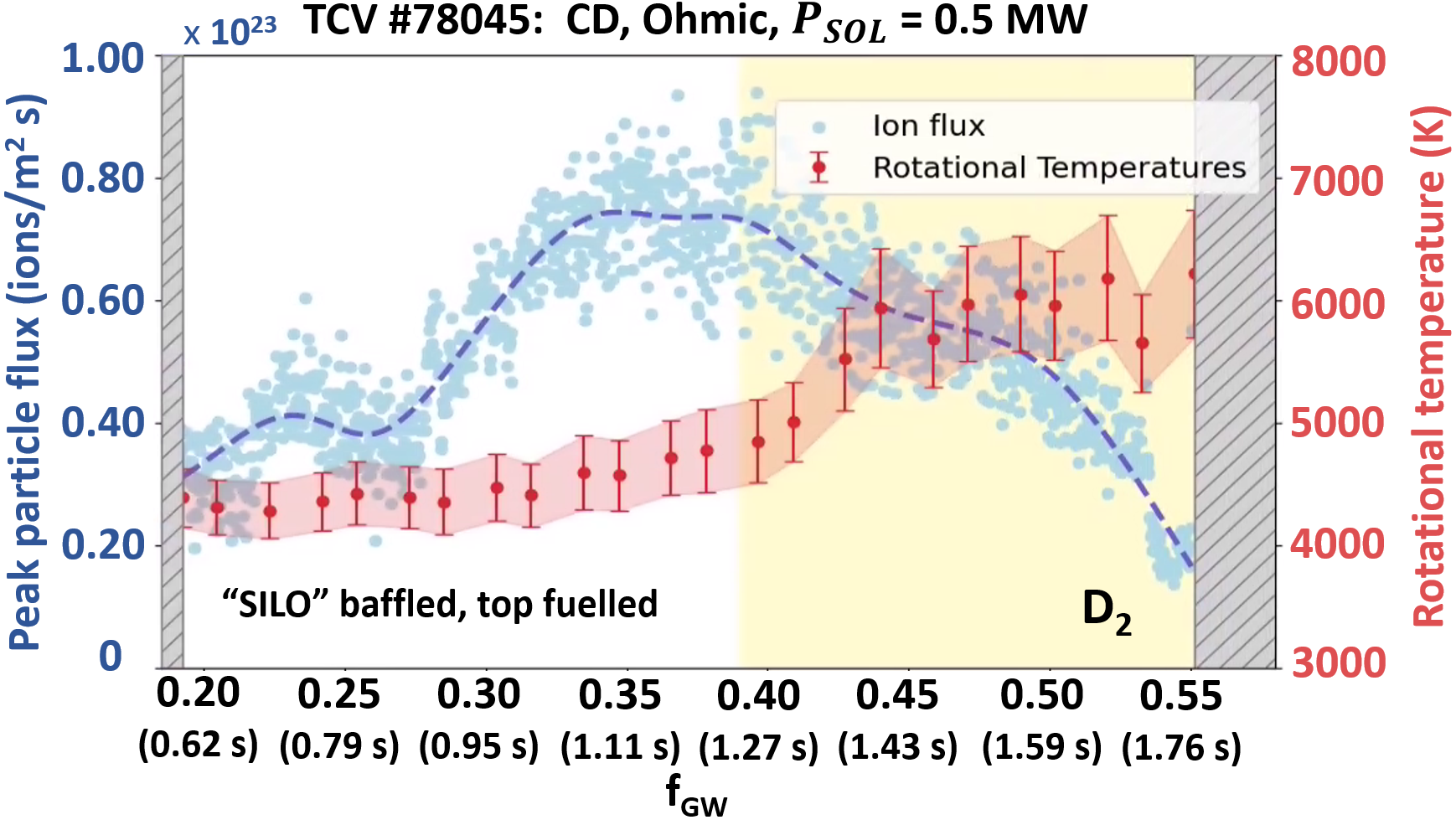}
            \put(15,47){\small \textbf{(b)}}
        \end{overpic}
    \end{minipage}
    \hspace{0.02\textwidth}
    \begin{minipage}[b]{0.47\textwidth}
        \centering
        \begin{overpic}[width=1.05\textwidth]{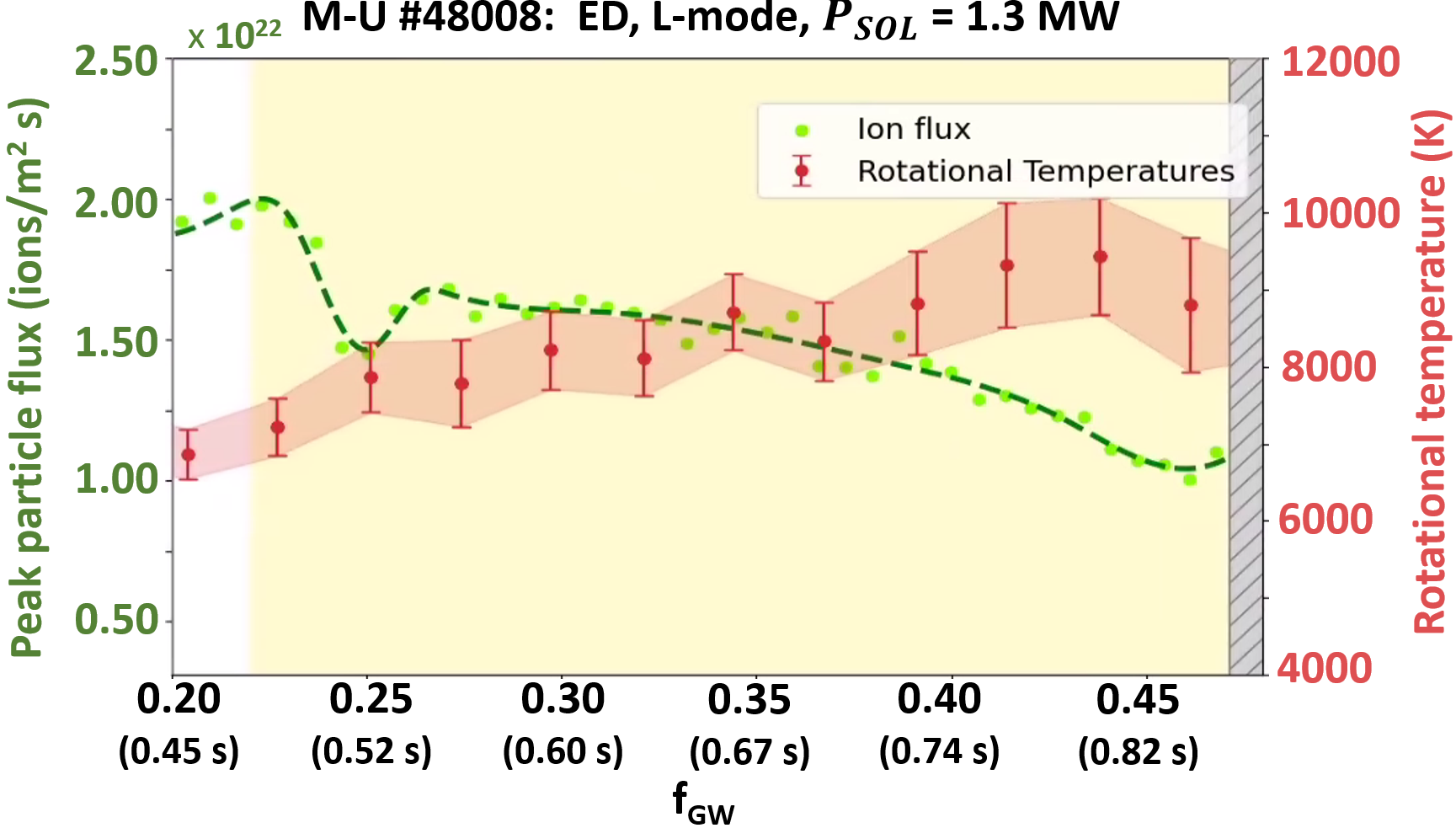}
            \put(15,47){\small \textbf{(f)}}
        \end{overpic}
    \end{minipage}
    \vspace{1em}

    \begin{minipage}[b]{0.47\textwidth}
        \centering
        \begin{overpic}[width=1.05\textwidth]{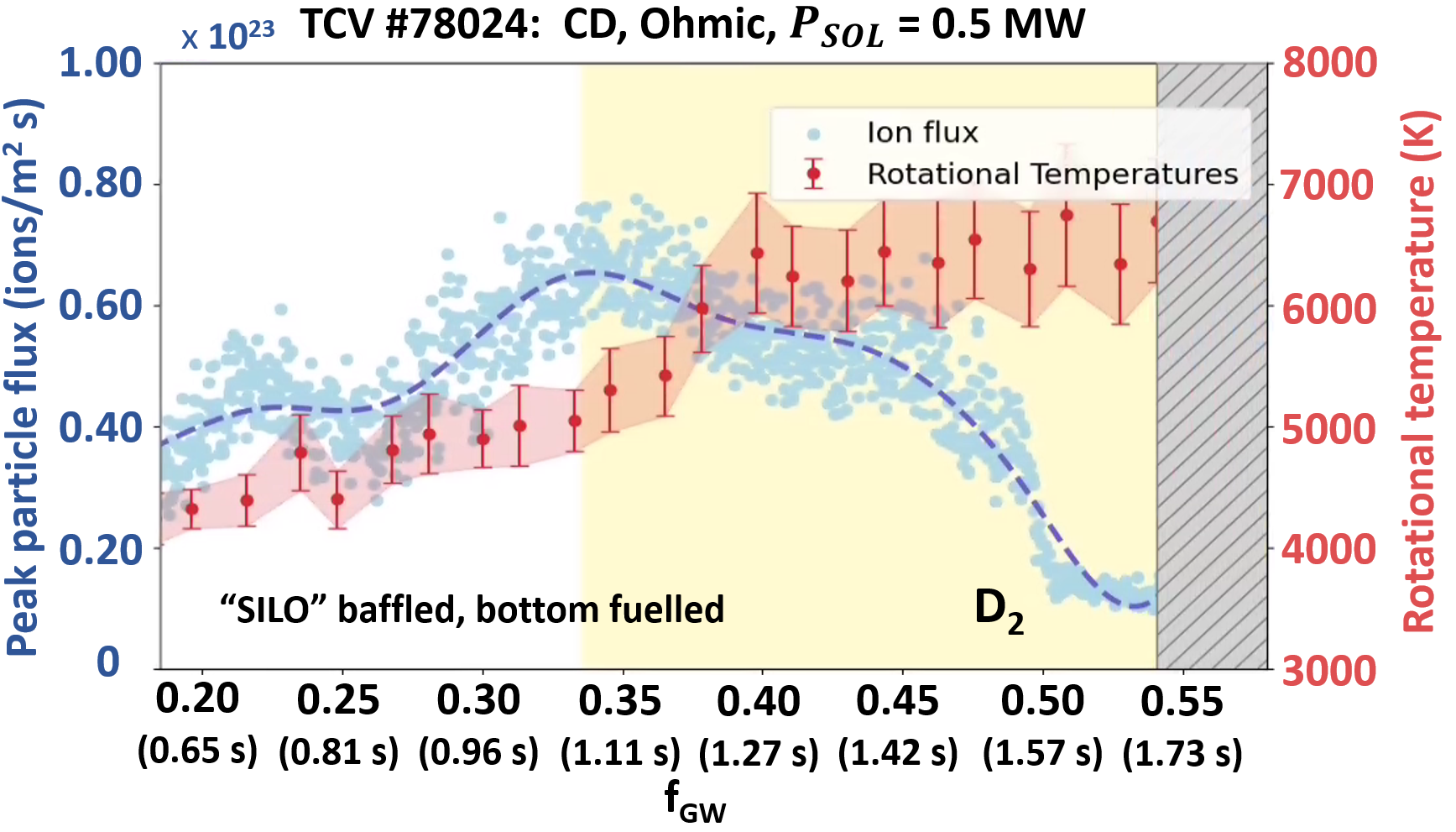}
            \put(15,47){\small \textbf{(c)}}
        \end{overpic}
    \end{minipage}
    \hspace{0.02\textwidth}
    \begin{minipage}[b]{0.47\textwidth}
        \centering
        \begin{overpic}[width=1.05\textwidth]{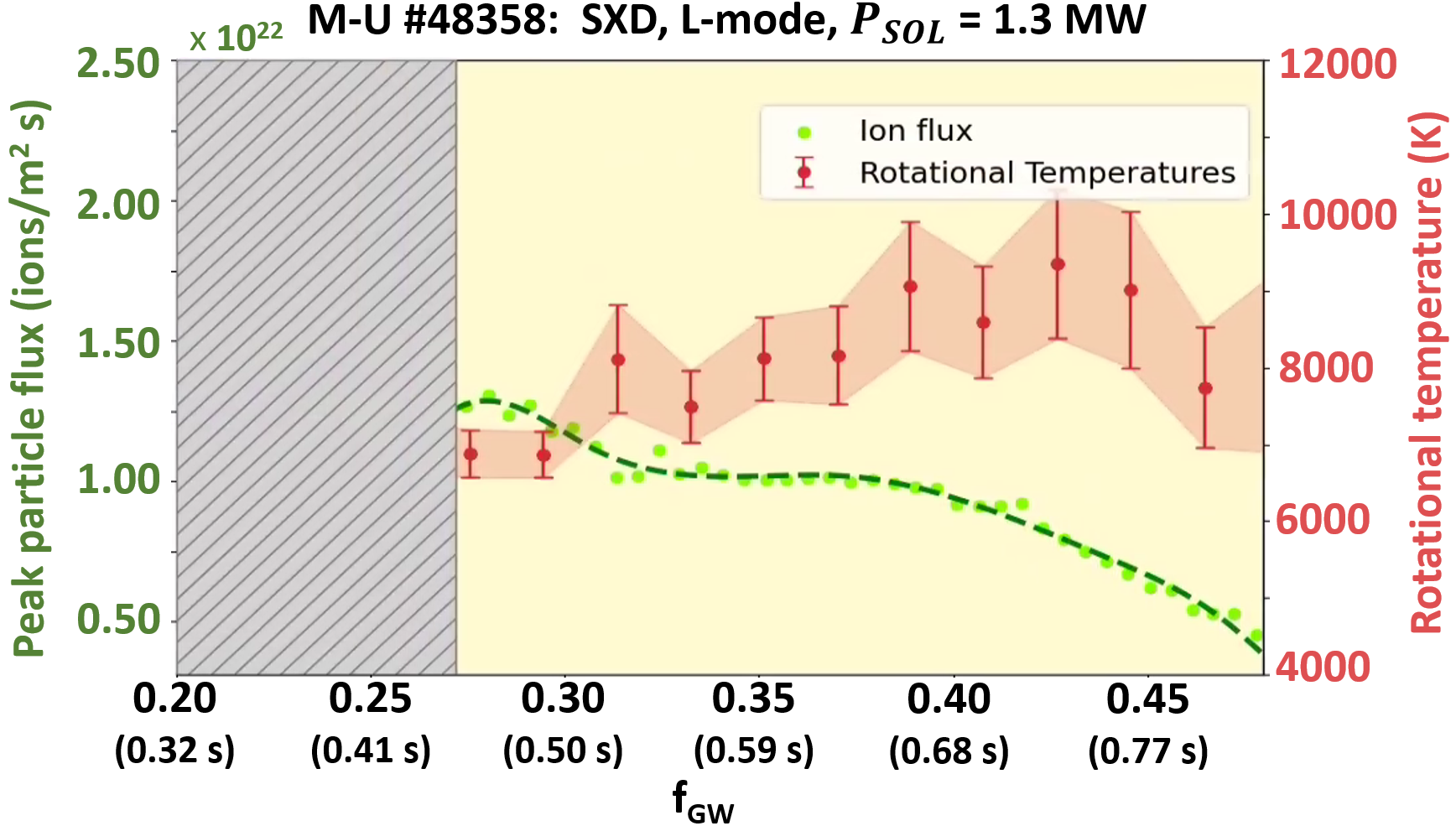}
            \put(15,47){\small \textbf{(g)}}
        \end{overpic}
    \end{minipage}
    \vspace{1em}

    \begin{minipage}[b]{0.47\textwidth}
        \centering
        \begin{overpic}[width=1.05\textwidth]{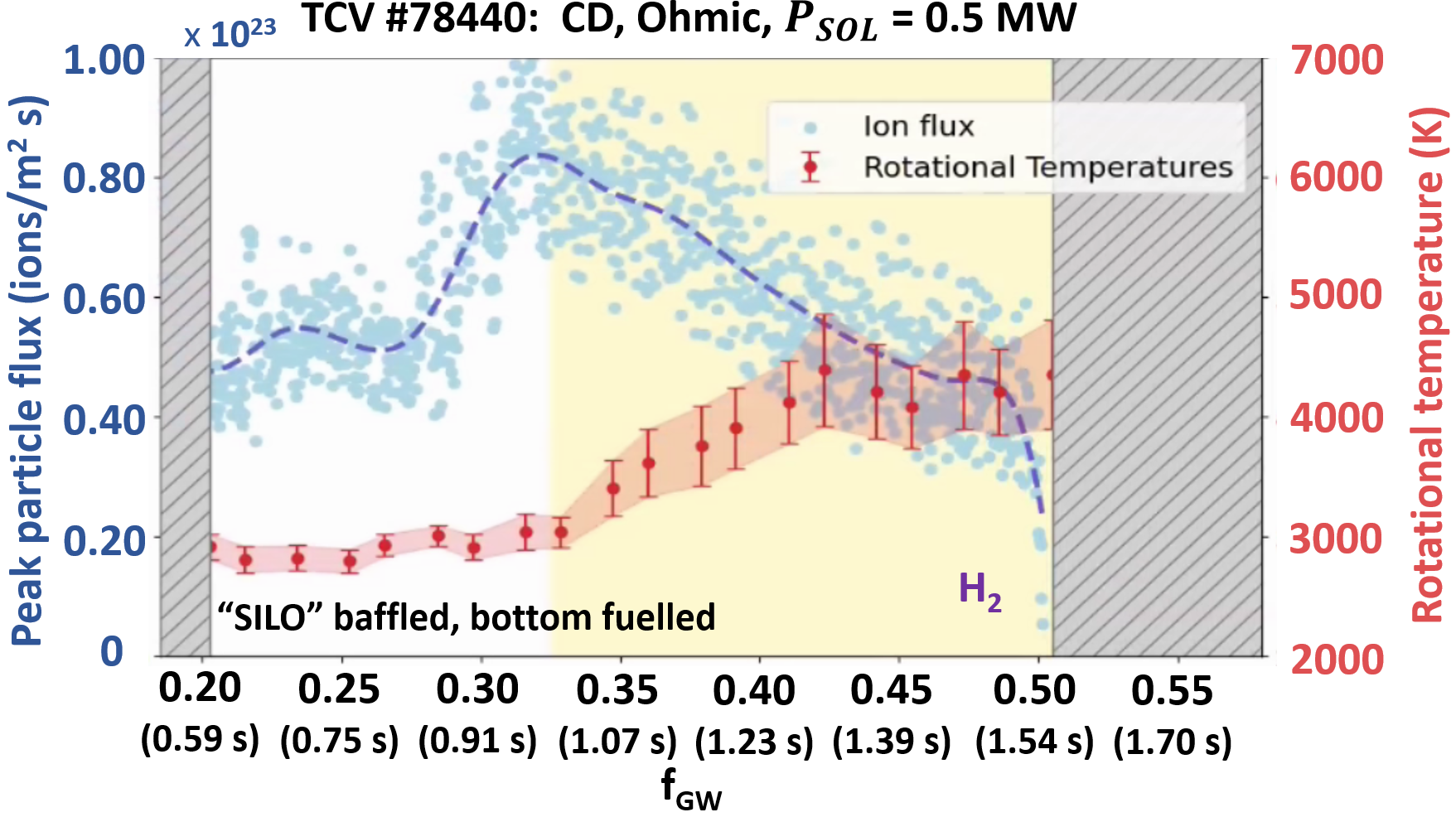}
            \put(15,47){\small \textbf{(d)}}
        \end{overpic}
    \end{minipage}
    \hspace{0.02\textwidth}
    \begin{minipage}[b]{0.47\textwidth}
        \hfill 
    \end{minipage}
    
    \caption{Graphs showing the evolution of rotational temperature at the strike point in several density ramps.  Plots (a), (b), (c) and (d) show Ohmic density ramps for TCV (note that (d) is an isotopic variation).  Meanwhile, plots (e) to (g) show CD, ED and SXD beam-heated L-mode discharges for MAST-U.  Note the different temperature scales.  Both Greenwald fraction and time are inidicated on the x-axis.  The lemon-coloured shading indicates the transition to detachment.}
    \label{SP_graphs}
\end{figure}

Generally, we observe that the rotational temperature of the molecules rises as the core density is increased and the divertor is cooled (figure \ref{SP_graphs}). The increase in rotational temperature occurs as the divertor nears detachment and increases during detachment, as can be seen from the saturation/decrease in ion target flux from where the graphs are shaded for ease of reference (although some increase in rotational temperature can occur before detachment). At the deepest levels of detachment, a decay in rotational temperature is occasionally observed (MAST-U ED, SXD ($f_{GW}>0.45$, figures \ref{SP_graphs}f,g)), which can be explained by the low plasma temperatures observed in these conditions ($T_e<0.3$ eV estimated in \cite{Verhaegh2023}), according to our reduced model (section \ref{sec6}). \\

Remarkably, similar qualitative behaviour is observed for all 5 variations for both MAST-U and TCV. Although neutral baffling results in higher rotational temperatures at the same $f_{GW}$ (figures \ref{SP_graphs}a,c), it also reduces the detachment onset density: the rotational temperature increases as detachment is approached and becomes deeper in both cases. The same holds true for a change in fuelling location: divertor fuelling, as opposed to main chamber fuelling (to regulate the core density), reduces the detachment onset density and increases the depth of detachment, leading to higher rotational temperatures (figures \ref{SP_graphs}b,c). Similar results were also obtained in MAST-U Ohmic conditions \cite{Osborne2023}, indicating that this qualitative behaviour (at the same level of detachment) is mostly independent of the level of heating applied. Even with $\mathrm{H}_2$, as opposed to $\mathrm{D}_2$, similar behaviour is observed (figures \ref{SP_graphs}c,d).\\

There are quantitative differences between the different cases, however. The rotational temperatures obtained for $\mathrm{H}_2$ are lower than for $\mathrm{D}_2$. The rotational temperatures obtained on MAST-U, as well as the range of its increase, are generally higher than on TCV. This may be associated with the strong baffling on MAST-U. \\

\begin{figure}
    \centering
    \begin{minipage}[b]{0.47\textwidth}
        \centering
        \begin{overpic}[width=1.05\textwidth]{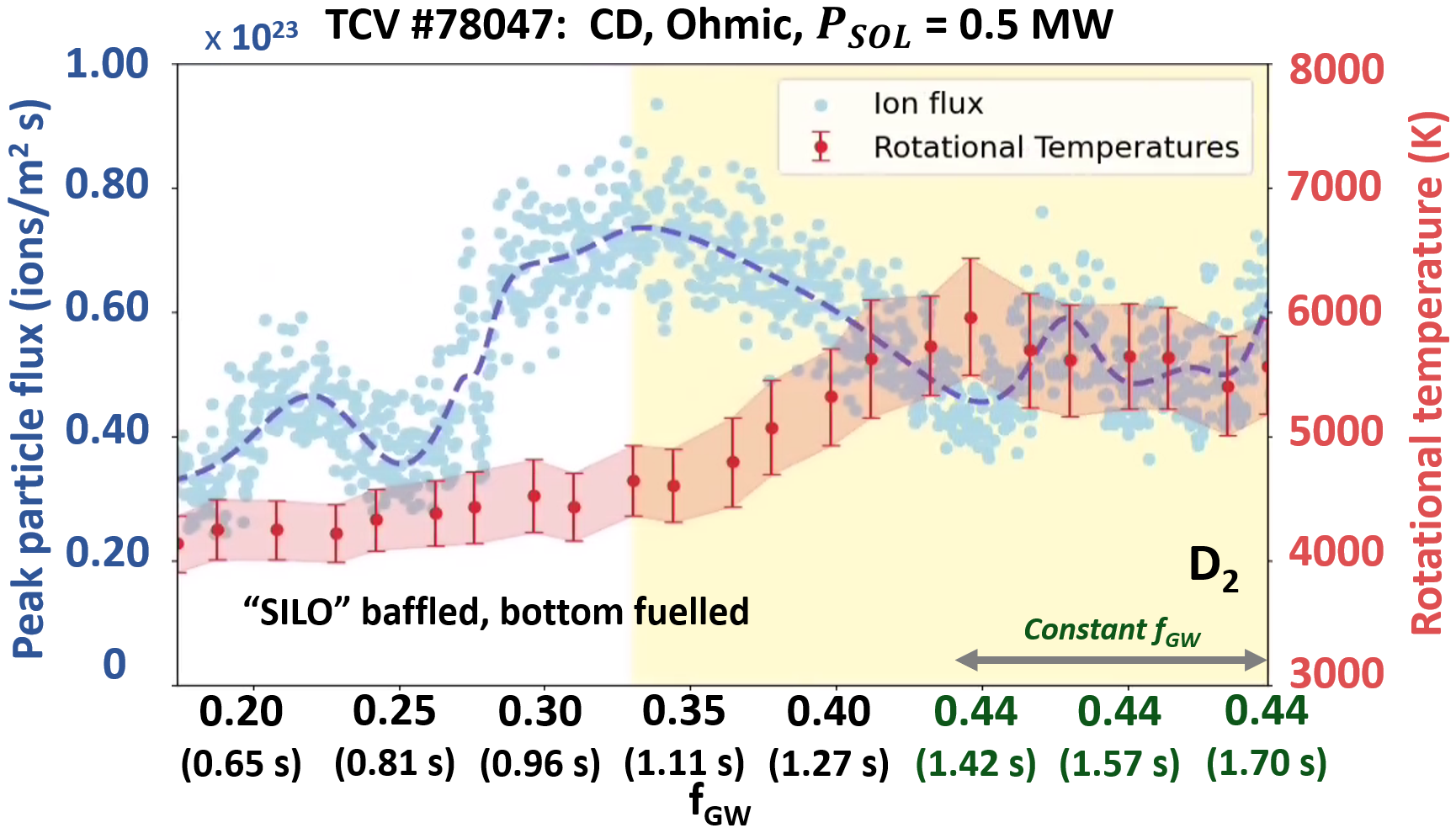}
            \put(15,47){\small \textbf{(a)}}
        \end{overpic}
    \end{minipage}
    \hspace{0.02\textwidth}
    \begin{minipage}[b]{0.47\textwidth}
        \centering
        \begin{overpic}[width=1.05\textwidth]{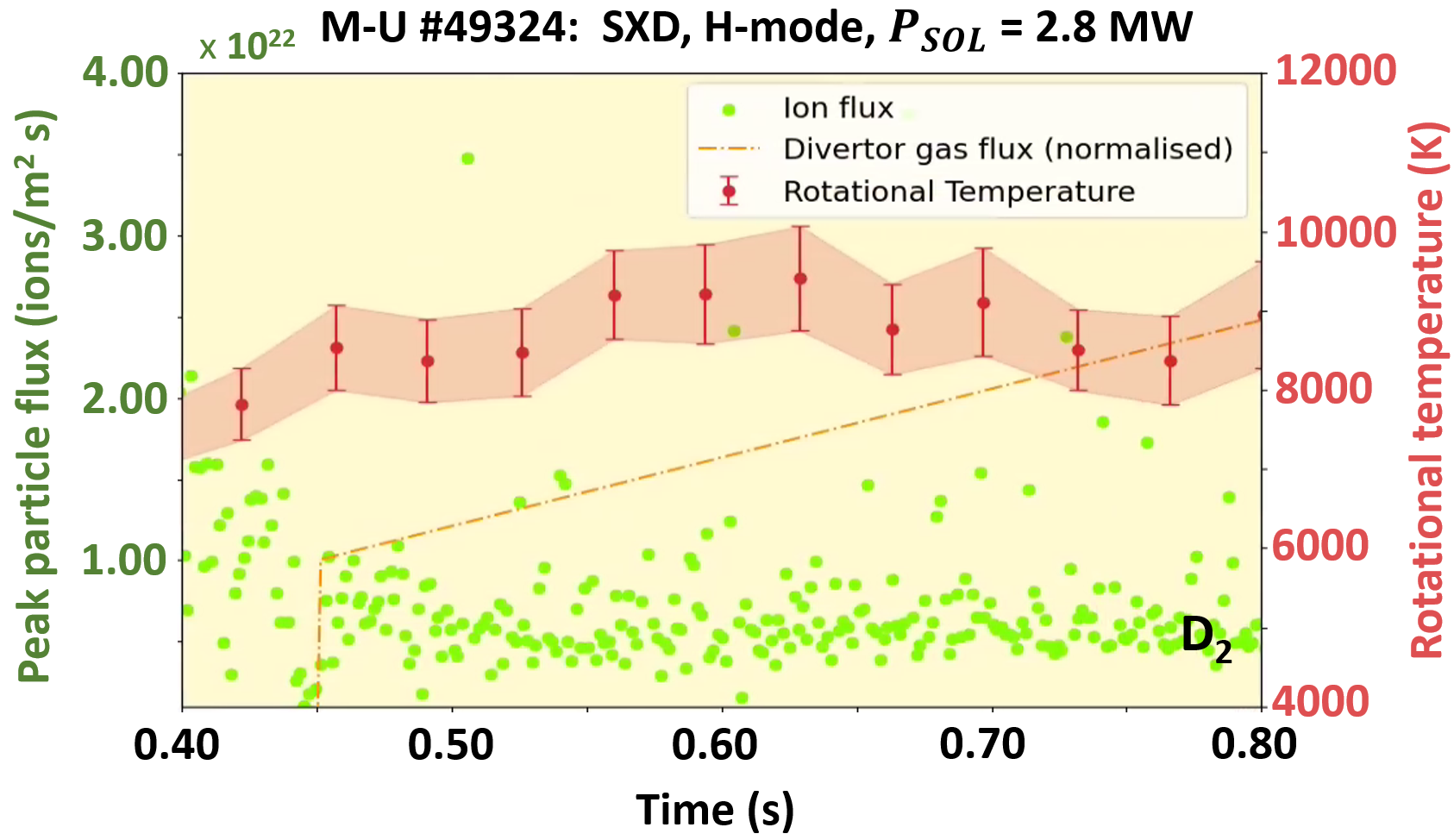}
            \put(15,47){\small \textbf{(b)}}
        \end{overpic}
    \end{minipage}

    \caption{Graphs showing the evolution of rotational temperature at the strike point in: (a) a TCV discharge with core density held constant from 1.4 s after initial ramp; (b) H-mode fuelling scan MAST-U.}
    \label{further_SP_graphs}
\end{figure}

Figure \ref{further_SP_graphs}a shows a TCV discharge in which the core density was held constant after an initial ramp.  We see that the rotational temperature increases and then holds roughly level with the core density indicating that divertor conditions hold steady with core density in such Ohmic discharges.\\

In MAST-U H-mode (higher heating power), where core density is kept constant, and divertor neutral pressure is increased through divertor fuelling (figure \ref{further_SP_graphs}b), a high rotational temperature with some increase is observed in keeping with the degree of detachment.\\

To obtain a clearer picture of the impact of divertor fuelling (room temperature $D_2$ from valve 3t (see figure \ref{MU_schematic})) on plasma-molecular kinetics and detachment, two additional MAST-U experiments (beam-heated L-mode) are performed. Both discharges are detached before divertor fuelling is introduced. During discharge 49286, two sharp $\mathrm{D}_2$ puffs were injected into the upper divertor from valve 3t (see figure \ref{MU_schematic}) at 400 ms and 600 ms (and two puffs into the lower divertor at 500 ms and 700 ms).  Discharge 49289 experienced a fuelling ramp, symmetric between lower and upper divertor. \\

\begin{figure}[h!]
    \centering
    \includegraphics[width=\textwidth]{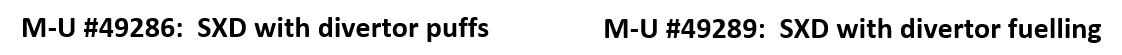}\\[1ex]
    
    \begin{subfigure}[b]{0.49\textwidth}
        \centering
        \includegraphics[width=\textwidth]{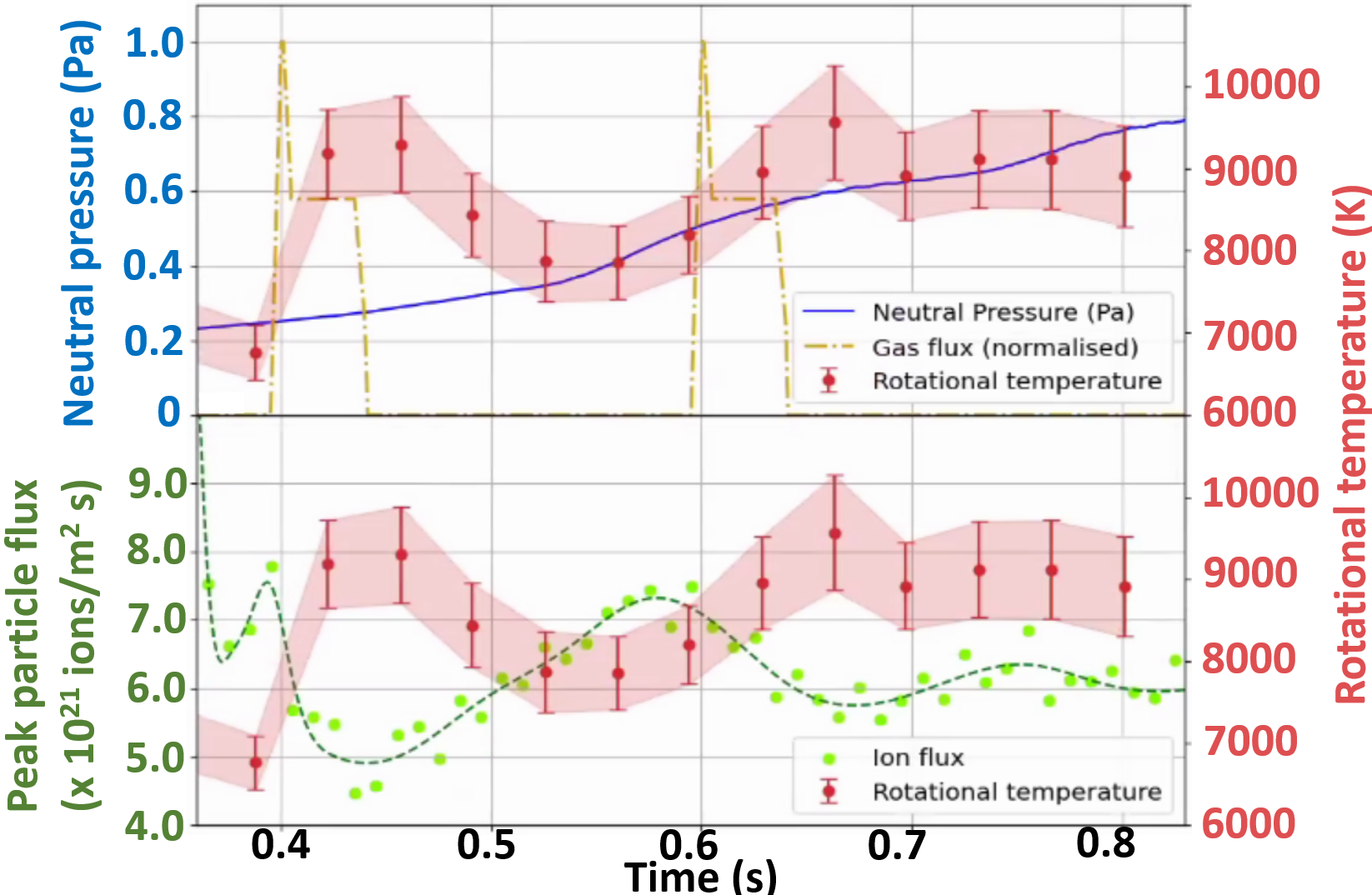}
        \caption{}
        \label{fig:np_49286}
    \end{subfigure}
    \hfill
    \begin{subfigure}[b]{0.49\textwidth}
        \centering
        \includegraphics[width=\textwidth]{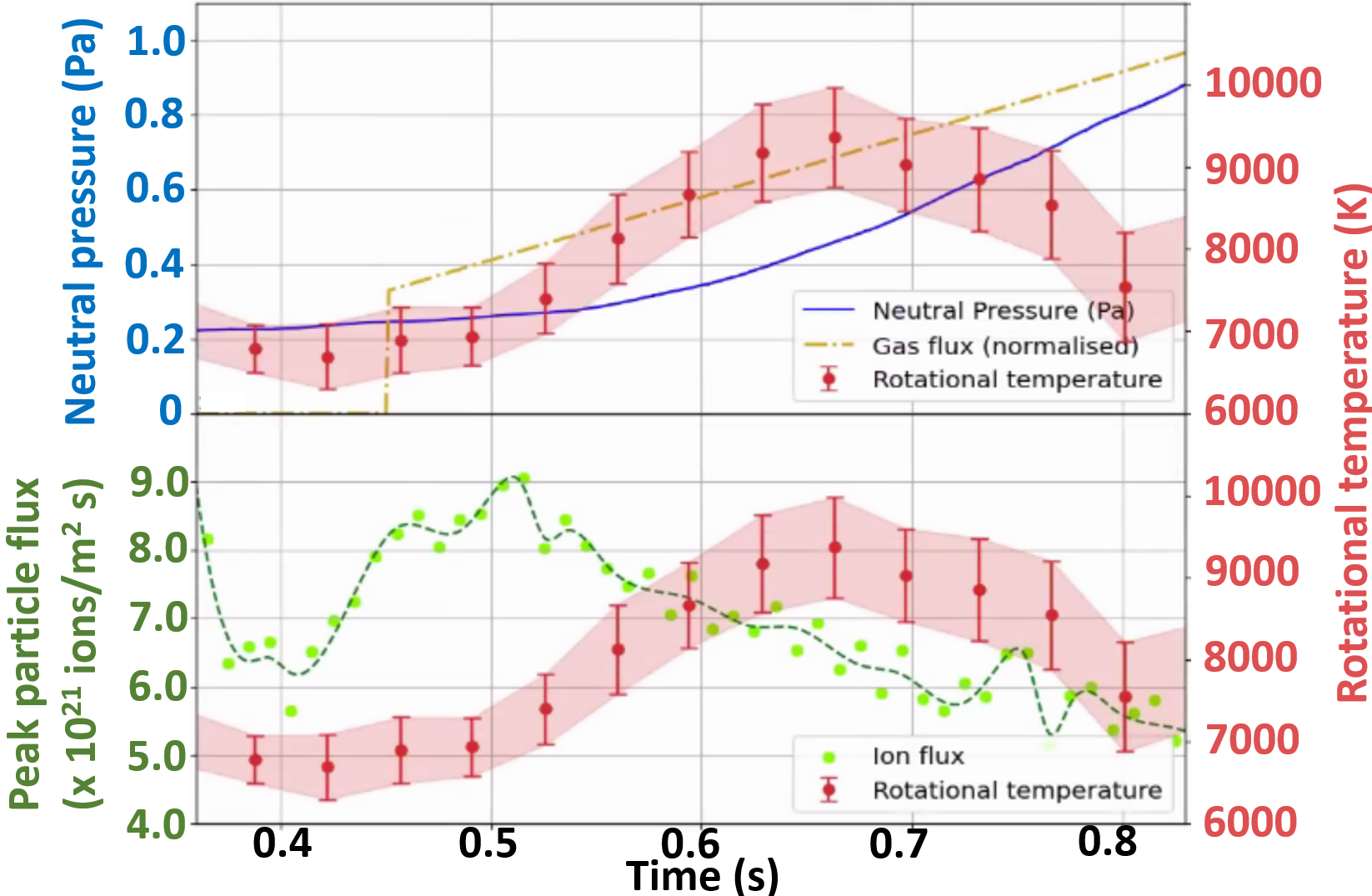}
        \caption{}
        \label{fig:np_49289}
    \end{subfigure}
    
    \caption{MAST-U discharges with divertor fuelling from valve 3t (see figure \ref{MU_schematic}). The top plots show the divertor $\mathrm{D}_2$ gas flux and the neutral pressure; and the lower plots show the peak particle flux at the strike point. The rotational temperature near the strike point is shown in both cases for reference. (a) shows 49286 in which two sharp puffs of $\mathrm{D}_2$ were injected at 400 ms and 600 ms. (b) shows 49289 during which there is a fuelling ramp.}
    \label{fig:np_comparison}
\end{figure}

For both cases, an \emph{increase} in rotational temperature (far beyond room temperature) is observed during \emph{room temperature} $\mathrm{D}_2$ gas injection in the divertor (figure \ref{fig:np_comparison}). The additional divertor cooling/deeper detachment induced by divertor fuelling (as evidenced by a reduction in ion target flux) results in an increase in rotational temperature. This observation is the most clear in the case where two sharp divertor puffs are introduced (figure \ref{fig:np_49286}), corresponding to an immediate reduction in ion flux and surge in rotational temperature, after which the ion flux and rotational temperature recovers until the second puff is introduced. Additionally, the upper divertor neutral pressure and rotational temperature do not respond to the staggered lower divertor puffs (at 500 ms and 700 ms) suggesting a decoupling between lower and upper divertor.  This is in keeping with finding by \citeauthor{BobK} \cite{BobK}.\\

To conclude, the rotational temperature rises throughout the detachment process. The increase tends to be more pronounced when particle flux ``rolls over" and the plasma becomes deeply detached.\\

\subsection{Vibrational distribution}

Using $\mathrm{D}_2$ Fulcher emission, the vibrational distribution of the first four vibrational bands (in the Fulcher state - $d^3 \Pi_u$) is measured at the strike point location. The vibrational distribution determines, to a large extent, the chemical interactions that the molecules can undergo with the plasma that result in the formation of molecular ions that lead to MAR and MAD.\\

\begin{figure}[h]
    \centering
    \begin{subfigure}[b]{0.44\textwidth}
        \centering
        \includegraphics[width=\textwidth]{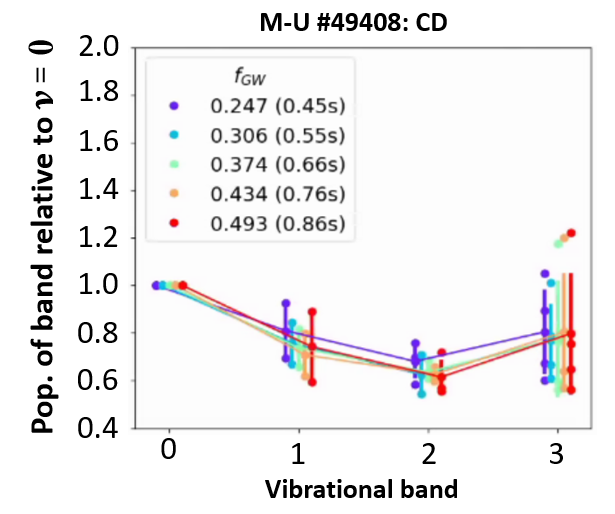}
        \caption{}
        \label{fig:a}
    \end{subfigure}
    \hfill
    \begin{subfigure}[b]{0.46\textwidth}
        \centering
        \includegraphics[width=\textwidth]{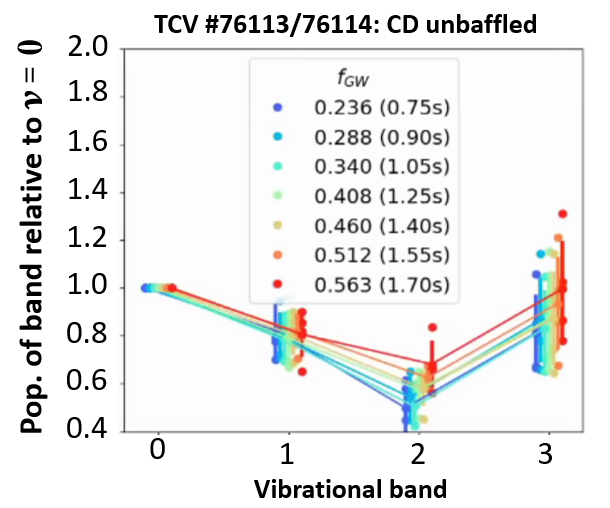}
        \caption{}
        \label{fig:b}
    \end{subfigure}
    \caption{The evolution of the vibrational distributions of bands $\nu=0,1,2,3$ for (a), a CD MAST-U density ramp, and (b), an unbaffled TCV density ramp.}
    \label{fig:VIB_cd_unbaff}
\end{figure}
\begin{figure}[ht]
  \centering
  \begin{subfigure}[b]{0.47\linewidth}
    \centering
    \includegraphics[width=\linewidth]{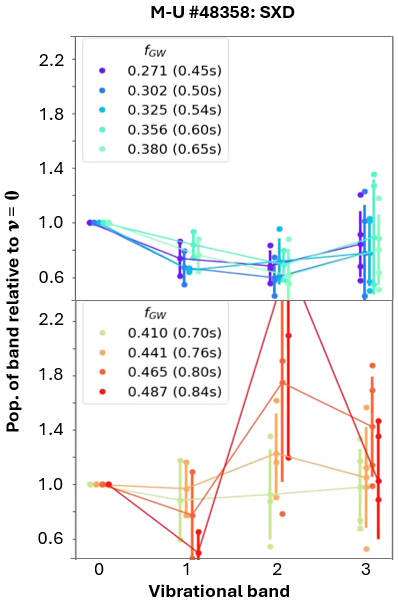}
    \caption{}
    \label{vib-dist-a}
  \end{subfigure}
  \hfill
  \begin{subfigure}[b]{0.47\linewidth}
    \centering
    \includegraphics[width=\linewidth]{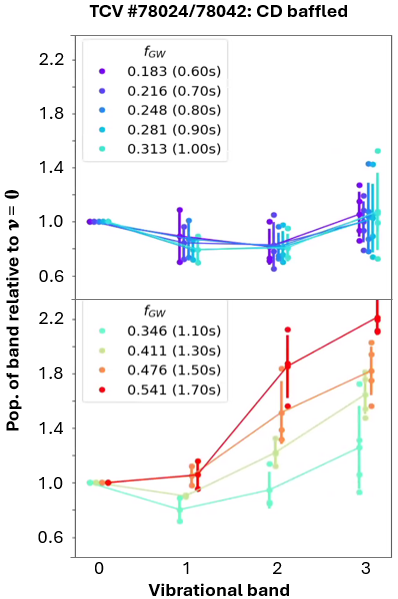}
    \caption{}
    \label{vib-dist-b}
  \end{subfigure}
  \caption{Graphs showing the vibrational distribution of the first four vibrational bands for an SXD density ramp MAST-U (a) and baffled density ramp TCV (b). The populations of bands $\nu$=1,2,3 are shown relative to $\nu$=0. The top and bottom graphs in each subfigure show the distribution before and after a population boost occurs for $\nu$=2,3.}
  \label{vib-dist}
\end{figure}

The vibrational distribution changes after detachment onset (figures \ref{fig:VIB_cd_unbaff}, \ref{vib-dist}) for a range of MAST-U and TCV conditions. This is consistent with previous MAST-U Ohmic observations  \cite{Osborne2023} where a correlation between the rotational temperature and an elevation of the $\nu=2,3$ Fulcher bands was observed. Such an elevation is in disagreement with a Boltzmann distribution in the ground state, mapped to the upper Fulcher state; the vibrational distribution cannot be quantified with a vibrational temperature.   \\

To estimate the populations of the vibrational bands, we use rotational lines which occur unimpeded in the $\nu=0$ band and in at least one more to build a composite plot with as much information as possible that can then be averaged. The full technique is described in the Appendix of \cite{Osborne2023}. In MAST-U experiments, this was possible in one single discharge since the ``DIBS" spectrometer (Divertor Imaging Balmer Spectrometer) has sufficient coverage (595-626 nm) to capture all four bands at high resolution (0.04 nm) \cite{Osborne2023}.  However, in TCV experiments, a repeat discharge was required to capture the information.\\

In cases where detachment does not occur (or is very weak) (figures \ref{SP_graphs}a,e) the vibrational distribution (relative to $\nu=0$ in the Fulcher state) changes negligibly (figures \ref{fig:VIB_cd_unbaff}a MAST-U, CD; \ref{fig:VIB_cd_unbaff}b TCV, unbaffled).  This is in contrast to the strongly detached TCV baffled and MAST-U SXD discharges, where a clear change in the vibrational distribution during the core density ramp is observed. To allow ease of reference, these two different density ranges have been separated in figure \ref{vib-dist}. The vibrational distribution starts to change near (TCV) or after (MAST-U) the detachment onset (figures \ref{SP_graphs}b and \ref{SP_graphs}g respectively) and becomes larger as detachment progresses.  \\

\section{Discussion}\label{sec6}

We have the following two principal observations from the experimental results:
\begin{itemize}
    \item The rotational temperature of the molecules in the divertor is seen to rise during the detachment process, even when the introduction of \emph{cold} molecules in the divertor further deepens detachment.
    \item Deepening of detachment results in a change of vibrational distribution in both MAST-U and TCV, which cannot be quantified with a Boltzmann distribution and a vibrational temperature in the ground state \cite{Osborne2023}.
\end{itemize}

\subsection{Understanding the increased gas temperature during detachment}

As was discussed earlier, the rotational temperature of the molecules in the ground state of the first vibrational band, is often taken as a proxy for the $\mathrm{D_2}$ gas temperature \cite{Trot-Tgas-Astashkevich}. We aim to explain why this gas temperature increases during detachment using a reduced model and exhaust modelling comparisons.

\subsubsection{Reduced model for molecular gas temperature} \label{sec6}

To explain the increase in rotational, or `gas' temperature, we imagine a reduced model describing neutrals ($\mathrm{D}$ and $\mathrm{D}_2$) in a bath of $\mathrm{D}^+$ ions at temperature $T_i$ undergoing elastic collisions with the ions. \\

This reduced model is based on the following assumptions/principles, resulting in the conclusion that energy transfer from the ions to the molecules is more efficient at lower plasma temperatures, when the lifetime of the molecules in the divertor is increased.
\begin{itemize}
    \item The dominant energy transfer mechanism from the plasma to the molecules is assumed to be ion-molecule elastic collisions.
    \item The temperature reached by the molecules via this mechanism is limited by their lifetime in the region of interest, which ends when they are either destroyed through a plasma reaction, or transported out of the region.
\end{itemize}

We can calculate how the temperature $T$ of the $\mathrm{D}_2$ molecules evolves over time by considering the energy entering and exiting the molecular cloud in the plasma leg.  This leads to the following differential equation for the gas temperature (derived in the appendix):
\begin{equation}
    \frac{dT}{dt}\approx0.4\,n_i\,\sigma_{rep}\left[\frac{k(T+2T_i)}{m_i}\right]^{\frac{1}{2}}(f\,T_i-T)-\frac{T}{\tau_{eff}}+\frac{T_{wall}}{\tau_{eff}}.
    \tag{\ref{final_model_de}}
\end{equation}

In equation \ref{final_model_de}, the first term on the right-hand-side originates from energy being transferred to the molecular cloud due to ion-molecule elastic collisions, the second term on the right-hand-side accounts for losses due to either destruction or transport, and the third term replaces lost molecules at the wall temperature to conserve molecular density.  $n_i$ is the ion density, and $\sigma_{rep}$ is a representative cross-section for elastic collisions between ions and molecules at a given temperature extracted from the Amjuel Eirene database \cite{AMJUEL} --- see Appendix).  $f$ is an efficiency term to account for exchanged energy which does not end up in the translational or rotational energy of the molecules (for example energy causing vibrational excitation).  $\tau_{eff}$ encapsulates both the characteristic destruction time $\tau_{dest}$ and transit time $\tau_{tranp}$:
\begin{equation}
    \frac{1}{\tau_{eff}}=\frac{1}{\tau_{dest}}+\frac{1}{\tau_{transp}}.
\end{equation}
The characteristic destruction time $\tau_{dest}$ is expected to dominate ($\tau_{dest}<<\tau_{transp}$) in a hot plasma.  To calculate $\tau_{dest}$, the following destruction mechanisms are considered: molecular ionisation $e+\mathrm{D}_2\rightarrow 2e+\mathrm{D}_2^+$; electron-impact dissociation $\mathrm{e}^- + \mathrm{D}_2\rightarrow \mathrm{e}^- +\mathrm{D}+\mathrm{D}$; and molecular charge-exchange $\mathrm{D}^++\mathrm{D}_2\rightarrow \mathrm{D}+\mathrm{D}_2^+$.  Rates for these reactions are sensitive to plasma temperature and electron density \cite{AMJUEL}.\\

\begin{equation}
    \frac{1}{\tau_{dest}}=n_e{\langle\sigma v\rangle}_{ionisation}+n_e\langle\sigma v\rangle_{dissociation} + n_e\langle\sigma v\rangle_{CX}.
\end{equation}
\\
In the model, we assume $T_i=T_e$ and $n_e=n_i$.  $\tau_{destroy}$ is primarily driven by electron-impact dissociation in attached conditions ($T_e > 5$ eV).\\

The transport time $\tau_{transp}$ limits the transit time of a molecule across the divertor (e.g. plasma) leg.  It is calculated assuming a simple 1D random walk (see appendix for further details).\\

The transport time $\tau_{transp}$ is expected to dominate ($\tau_{transp}<<\tau_{dest}$) in a cold plasma.
\\
\\
For a given ion temperature, while $\sigma_{rep}$ varies as $T$ evolves, this variation is very small.  If we also take a characteristic $\tau_{transp}$ such that we neglect its variation with $T$, and use the approximation $T<<T_i$, we can write down an approximate equilibrium ($\frac{dT}{dt}=0$) analytical solution for equation \ref{final_model_de}:
\begin{equation}
    T\approx\frac{0.4\,f\,n_i\,\sigma_{rep}\,v_{th}\,T_i+\frac{T_{wall}}{\tau_{eff}}}{\frac{1}{\tau_{eff}}+0.4\,n_i\,\sigma_{rep}\,v_{th}},
\end{equation}
where $v_{th}=\sqrt{2kT_i/m_i}$.\\

\begin{figure}[h!]
    \centering
    \includegraphics[width=0.6\linewidth]{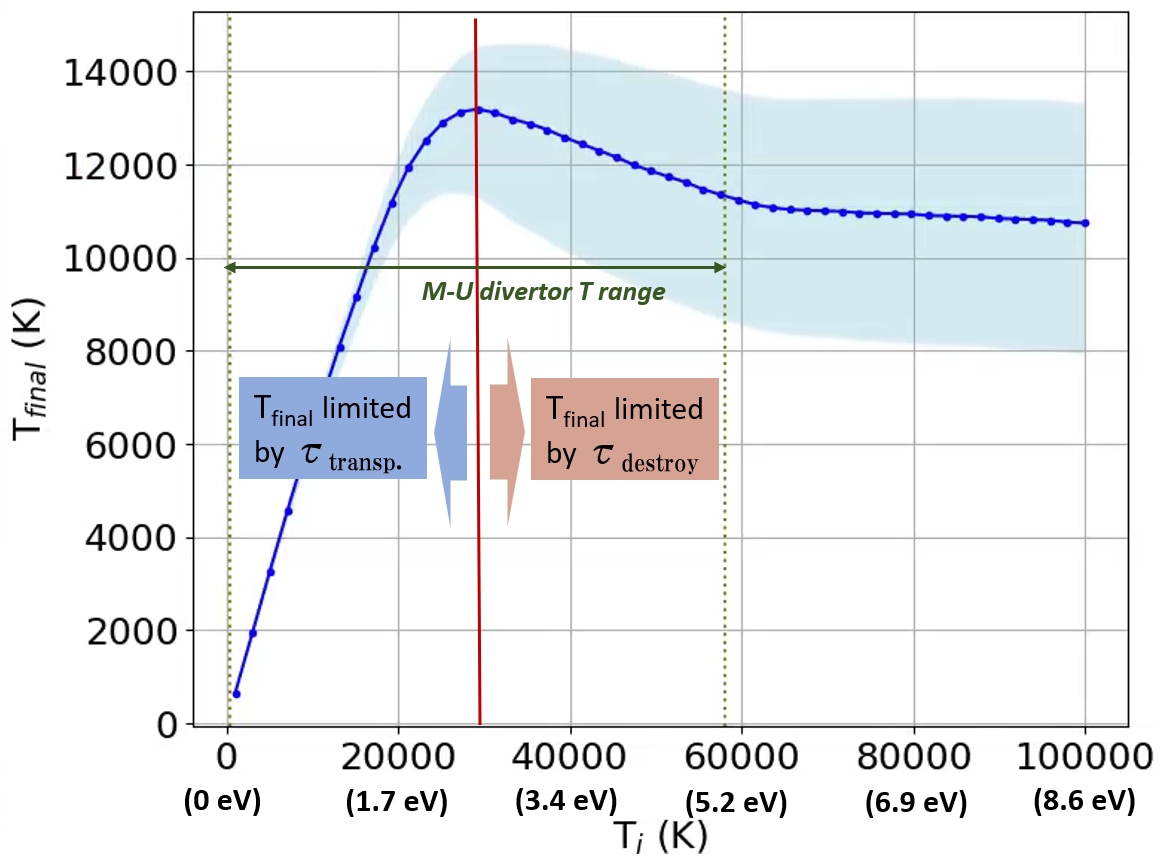}
    \caption{The final molecular temperatures achieved in the simplified model for a complete scan of ion-bath temperatures.  $n_i=2\times10^{19}\,\si{m^{-3}}$.  The shaded region allows for uncertainty in $\sigma$ of $30\%$. The dotted green vertical lines indicate the range of divertor temperatures in the MAST-U divertor during a typical SXD discharge.}
    \label{fig:T_final_full_scan}
\end{figure}

From this solution, using MAST-U relevant parameters, and assuming an energy exchange efficiency of $f=0.75$, we obtain figure \ref{fig:T_final_full_scan} showing the equilbrium gas temperature for various ion temperatures, taking into account the lifetime of the molecules due to plasma reactions, and their transit time through the plasma leg.\\

The highest gas temperature is reached at moderate ion temperatures of a few eV - cold enough to provide relatively long molecular lifetimes, but energetic enough to heat up the molecule within its transit time.  This is qualitatively consistent with the observation that the rotational temperature increases during detachment: the gas temperature increases because the molecules live longer and undergo more collisions with the plasma. Introducing fuelling into a detached divertor leads to deeper detachment and longer molecular lifetimes, increasing the gas temperature obtained. A decay of rotational temperature is observed in the deepest detached conditions (\ref{SP_graphs}f,g), where the plasma is insufficiently energetic to heat up the molecules (within their transit times), in agreement with the model. The range of gas temperatures obtained by the model are in reasonable quantitative agreement with the MAST-U measurements. \\

This model is intended to capture key physics in a (semi-)classical approach, and complications are avoided where reasonably possible.  We, therefore, note the following limitations of our model:
\begin{itemize}
    \item The use of a representative constant density.
    \item The use of a constant cross-section in the energy exchange rate from \cite{Cravath}.  This is addressed by selecting an appropriate representative value (see Appendix) and allowing for a generous uncertainty.
    \item Neglecting other energy transfer pathways, in particular vibrational excitation.  This is addressed by incorporating an efficiency factor $f$ in the energy exchange term.
    \item Not evolving the ion temperature.  This may marginally inflate the molecular temperature reached.
    \item Not accounting for electron temperature effects.  Elastic collisions with electrons were investigated and our modelling suggests that in relevant divertor conditions, the electrons supply less than 10\% of the energy required to explain the increase in temperature.
\end{itemize}

It does not treat the various rovibronic transitions that could result in rotational excitation. Such processes have been considered in collisional-radiative models, where it was suggested that electron collisions can have a significant impact on the rotational distribution, particularly in high temperature conditions ($T_e\sim$10 eV) \cite{Yoneda2023SpectroscopicTokamaks}. \\

\subsubsection{Comparison against molecular gas temperature in SOLPS-ITER simulations}\label{sec7}

Using interpretive exhaust simulations by \citeauthor{D-Moulton} (similar to those in \cite{D-Moulton} but with 1 MW input power) of the MAST-U experiments studied, we compare the molecular gas temperatures obtained against experiments and investigate the mechanisms resulting in an increase in molecular gas temperature. \\

\begin{figure}[htbp]
    \centering
    \includegraphics[width=\textwidth]{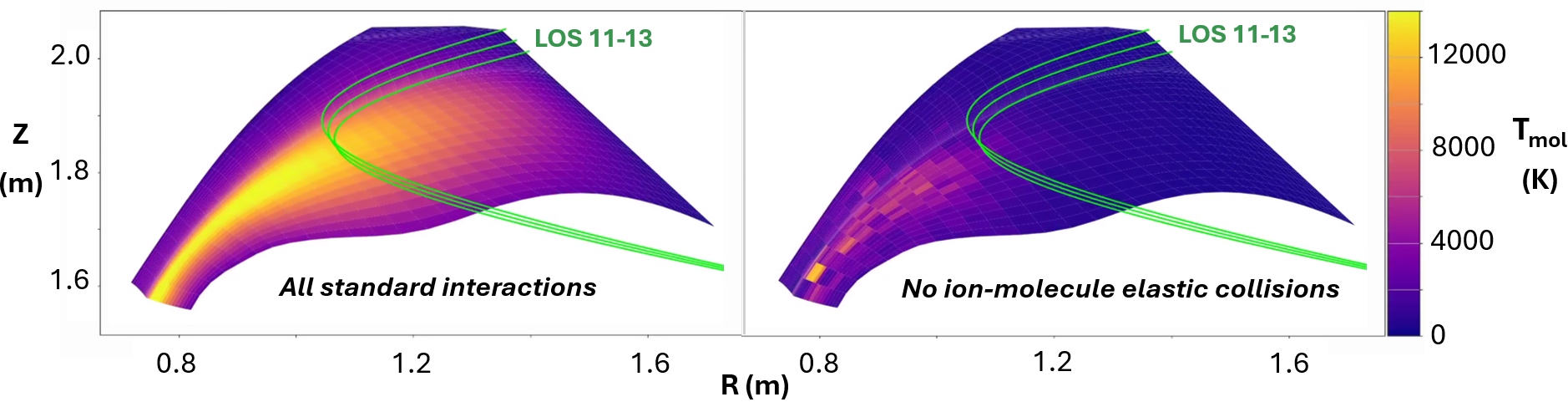}
    \caption{The molecular temperature generated by SOLPS-ITER (1 MW detached SXD) for all standard reactions (left plot) , and all reactions \textit{except} ion-molecule elastic collisions (right plot).}
    \label{fig:SOLPS_examples}
\end{figure}

The molecular gas temperature obtained from exhaust simulations is compared for a default rate setup and a modified setup where the only change is the disablement of elastic collisions between ions and molecules (figure \ref{fig:SOLPS_examples}). This shows: 1) the molecular temperature obtained is significantly higher than the `cold' gas injected and is in a similar range as the (rotational) temperatures observed on MAST-U; and 2) the molecular temperature is primarily driven by elastic ion-molecule collisions. Additionally, a decrease in molecular temperature near the target is observed, consistent with our reduced model (figure \ref{fig:T_final_full_scan}) and experiments at deepest detachment (figure \ref{fig:np_comparison}b). This provides additional evidence that ion-molecule collisions are dominant in the energy transfer from the plasma to the molecules.  (It is worth noting that the remaining high temperatures in the right-hand-side plot of figure \ref{fig:SOLPS_examples} are due to atom-molecule collisions, and that the molecular temperature flattens completely if these are also disabled.)\\

To enable a quantitative comparison between experiments and simulations, a synthetic diagnostic setup is used. Here, the Fulcher-emission weighted averaged $\mathrm{D}_2$ temperature, along a spectroscopic line-of-sight is obtained. Since $\mathrm{D}_2$ Fulcher emission is used to infer the rotational temperature, the rotational temperature estimate corresponds to the region along the line-of-sight where emission occurs.\\

\begin{figure}[htbp]
    \centering
    \includegraphics[width=0.6\linewidth]{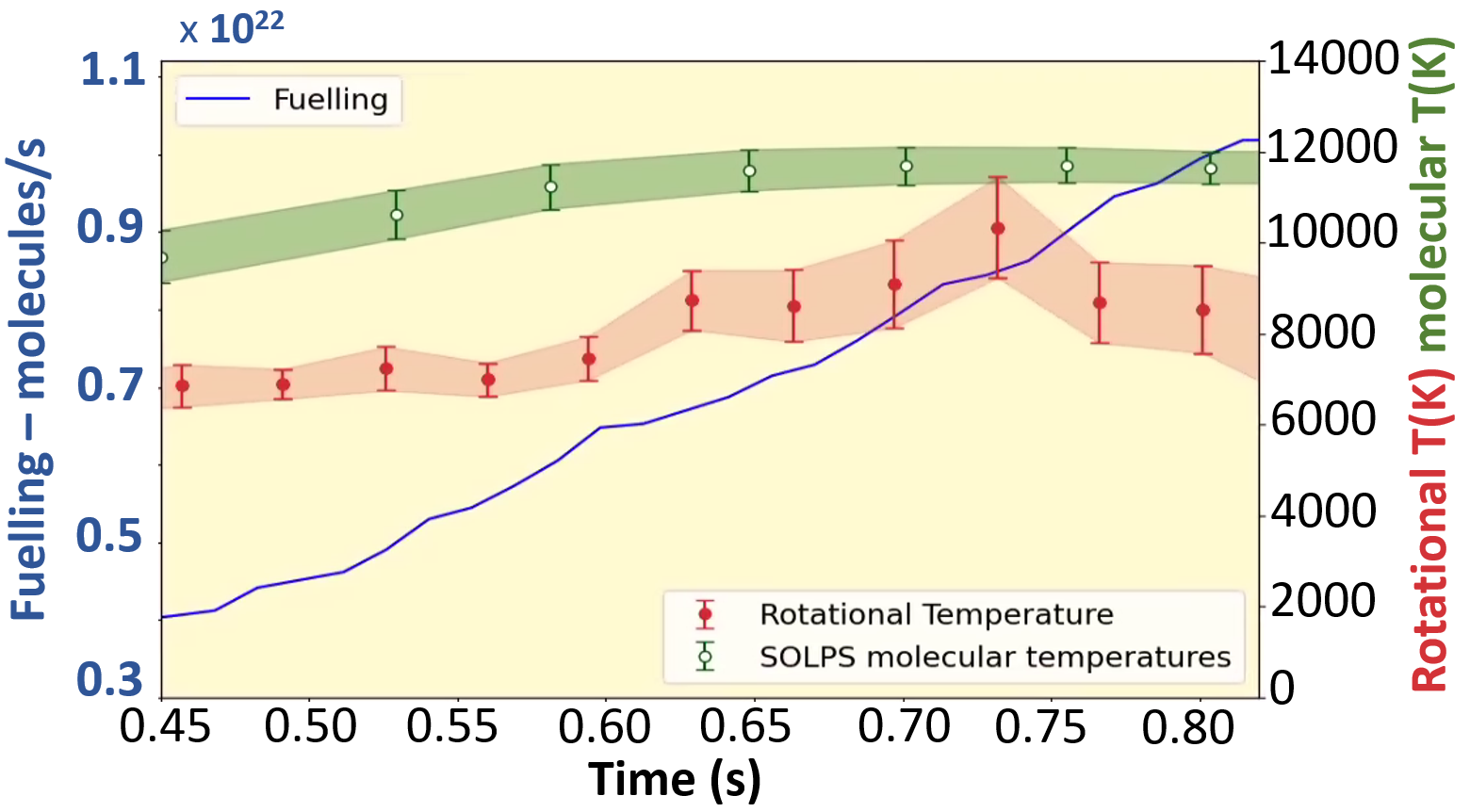}
    \caption{The red plot shows the experimentally determined rotational temperatures averaged across a group of LOS (11-13 shown on figure \ref{fig:SOLPS_examples})  at various degrees of detachment.  The green plot shows the SOLPS inferred molecular temperatures from several different simulations for the same LOS grouping.  The simulations are matched to the experiment via the fuelling rate (blue line).}
    \label{fig:Temps_vs_SOLPS}
\end{figure}

We compare an SXD density ramp (discharge 48358) against a sequence of simulations for an SXD configuration with very similar geometry.  Figure \ref{fig:Temps_vs_SOLPS} shows the result.  The correspondence is made via the fuelling rate (blue line) - but some caution is needed noting that, although the fuelling location is the same in both cases, there is considerable uncertainty in the fuelling rate experimentally speaking.  We also note that the input power is 1 MW in the simulations, but has considerable uncertainy in the experimental discharge ($1.3\,\pm0.2\,\si{MW}$).  Despite these caveats, the synthetically obtained molecular temperatures in the SXD simulations have a range similar to the rotational temperatures experimentally observed.  The increase in temperature with detachment is also in reasonable agreement.

\subsection{Vibrational distribution}

To obtain further insight into the expected vibrational distribution in the $\mathrm{D}_2$ Fulcher state, the vibrational distribution in the electronic ground state ($X^1 \Sigma_g$) is modelled using a collisional-radiative model setup \cite{Verhaegh2024} that mimics the Eirene vibrationally resolved setup (H2vibr) \cite{Wiesen2015} with corrected molecular charge exchange rates \cite{Ichihara2000} using the CRUMPET \cite{Holm-crumpet} framework (see \cite{Verhaegh2024, Stijn} for more details). With this, we can model the $X^1 \Sigma_g$ vibrational distribution that was implicitly assumed in the derivation of the rates employed by Eirene. Franck-Condon coefficients \cite{Fantz1998SpectroscopicMolecules, Briefi2020APlasmas} are used to map the $X^1 \Sigma_g$ vibrational distribution to the $\mathrm{D}_2$ Fulcher state ($d^3 \Pi_u$), using the approach highlighted in \cite{Osborne2023}. It should be noted that both a setup identical to Eirene and a modified setup with corrected molecular charge exchange rates (both presented in \cite{Verhaegh2024}) were used for this study and resulted in the same interpretation. \\

\begin{figure}[htbp]
    \centering
    \includegraphics[width=\linewidth]{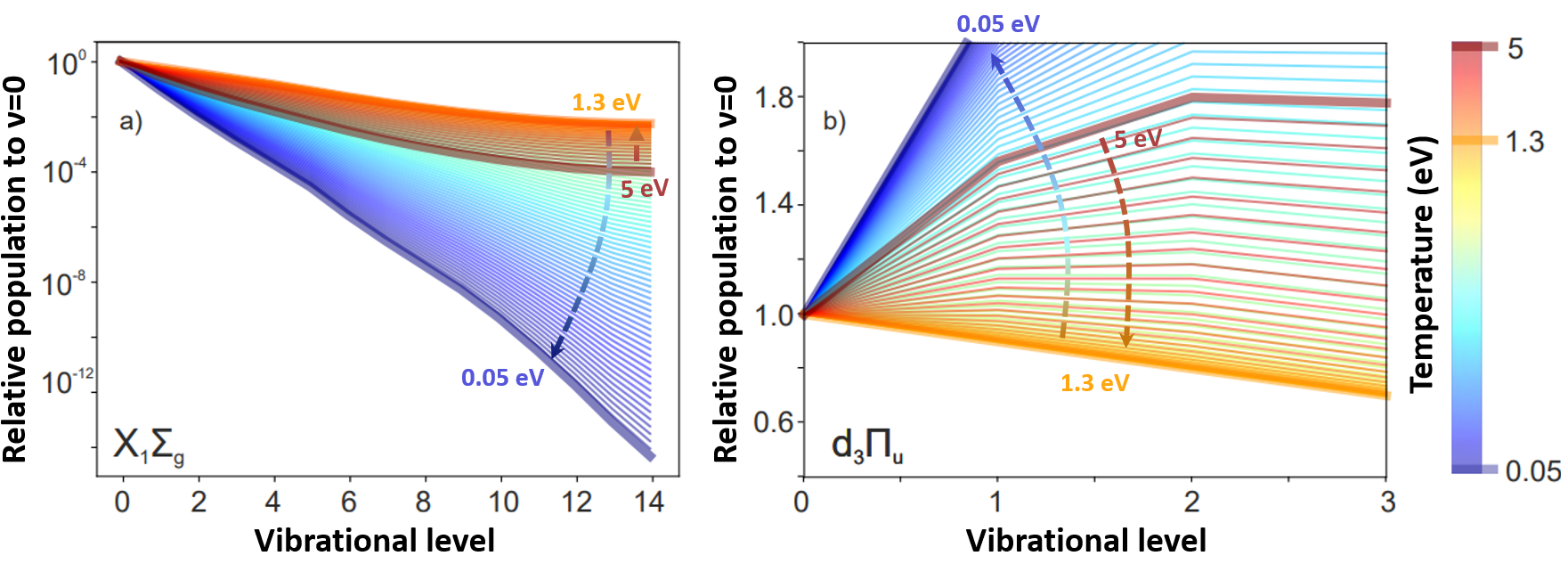}
    \caption{Vibrational distribution in the electronic ground state (a, $X_1\Sigma_g$) and upper Fulcher state (b, $d_3 \Pi_u$), relative to the $\nu=0$ population (for the respective electronic state) for different temperatures ranging from 5 eV to 0.05 eV according to an `Eirene-like' CRM setup. The distribution in the upper Fulcher state is obtained through mapping the ground state distribution using a Franck-Condon approach.}
    \label{fig:Vibr_Dist_Model}
\end{figure}

With this approach, we can compute the $d^3 \Pi_u$ $\nu=1,2,3$ distribution, relative to $d^3 \Pi_u$ $\nu=0$, as a function of $T_e$ (which is assumed to be equal to the ion temperature; a reasonable approximation according to SOLPS-ITER simulations). According to the modelled distribution, an increase in the $d^3 \Pi_u$ $\nu=2,3$ states is expected for $T<1.3$ eV (figure \ref{fig:Vibr_Dist_Model}b), which seems to be qualitatively consistent with the experimentally observed change during detachment (figure \ref{vib-dist}). Comparing the modelled distributions in the $X^1 \Sigma_g$ and $d^3 \Pi_u$ $\nu=1,2,3$ states suggests that this increase in the upper Fulcher state corresponds to a reduction in population of the high vibrational levels in the electronic ground state.  This reduction in the ground state distribution occurs due to the electron temperature becoming insufficient to drive vibrational excitation, and the depletion of highly vibrationally excited molecules through MAR and MAD.  The surprising correspondence to \textit{increased} populations in the upper Fulcher state comes about because of the the non-trivial way in which the ground state distribution maps to it using Franck-Condon factors.\\

However, we note two disagreements with the experiment: 1) $d^3 \Pi_u$ $\nu=1,2,3$, with respect to $d^3 \Pi_u$ $\nu=0$, is expected to decrease from $>5$ eV to lower temperatures (before detachment) in the modelled distribution. However, no changes in the vibrational distribution are observed. This may be explained by the omission of vibrational excitation through electronic excitation followed by radiative decay (e.g. $e^- + \mathrm{D}_2 (\nu_1) \rightarrow e^- + \mathrm{D}_2^*$, $\mathrm{D}_2^* \rightarrow \gamma + \mathrm{D}_2 (\nu_2)$). This under-predicts vibrational excitation above 1.5 eV \cite{Stijn,Chandra2023}, resulting in a decrease in vibrational excitation above 1.5 eV, which does not occur if such reactions are included in the CRM \cite{Stijn}. \\

A second disagreement is the observation that $d^3 \Pi_u$ $\nu=1$ increases together with $d^3 \Pi_u$ $\nu=2,3$ for $T<1$ eV in the modelled distribution, which does not occur experimentally. The cause of this is unclear. Potentially, plasma-surface interactions in combination with transport of vibrationally excited molecules could result in a reduced temperature sensitivity for the $d^3 \Pi_u$ $\nu=1$ state, requiring further study. 

\subsection{Power and momentum losses from elastic plasma-molecule collisions}\label{sec8}

Elastic collisions between the ions and the molecules, resulting in a molecular `gas' temperature increase that is observed, are also accompanied with significant plasma power and volumetric momentum losses according to SOLPS-ITER simulations, playing a significant role in detachment \cite{Moulton, Omkar}. \\

Reduced models focussing on the neutral buffer during detachment have been proposed in the past to study volumetric momentum losses from plasma-atom interactions \cite{Self-Ewald, Pitcher1999}, as well as for the energy required to ionise (burn through) the neutral buffer with regard to transients \cite{stuart-H-reduced-models}.\\

In this section, we use ``back-of-the-envelope arguments", in combination with rotational temperature observations and other experimental measurements (ionisation front position, neutral pressure) to experimentally estimate power and momentum losses from elastic ion-molecule collisions, which are compared against SOLPS-ITER simulations. \\

\subsubsection{Experimental power loss estimates from elastic plasma-molecule collisions with modelling comparison}\label{sec8}

Assuming the rotational temperature is a proxy for the gas temperature, our rotational temperature measurements can be used as a proxy for these plasma power losses. To achieve this, we make a simplified estimate of the power losses obtained experimentally. First, the power transfer per molecule should be estimated. Assuming the molecules heat up from room temperature to the observed rotational temperature within their transit time, the power transfer per molecule can be estimated as \\

\begin{equation}
P_{mol}^{1} = \frac{5k\Delta T_{rot}}{2\tau_{transport}},
\end{equation}
where $\Delta T_{rot} = T_{rot} - T_{wall}\approx T_{rot}$.\\

To obtain the total energy loss, this should be multiplied with the total amount of molecules in the system, which depends on the molecular density ($n_{mol}$) and the volume of the neutral gas buffer ($V$).

\begin{equation}
P_{mol} = P_{mol}^{1} \times n_{mol} \times V.
\end{equation}

The molecular density can be estimated from the measured divertor neutral pressure $p_{n}=n_{n}^{gauge}kT_{gauge}$, where $n_n^{gauge}$ is the neutral density at the gauge. This is measured by the fast-ion gauge in the sub-divertor beyond the wall.  We use a molecular flow model, with conservation of flux, to simulate a synthetic baratron (as previously done in \cite{Niemczewski1997, Wischmeier2005}) to account for the various pipework in between the fast-ion gauge and the tokamak to link the molecular density in the divertor to the molecular density at the gauge:
\begin{equation}
        n_{mol}^{div} \approx n_{mol}^{gauge} \sqrt{\frac{T_{gauge}}{T_{mol}}}.
\end{equation}
Assuming that, in the sub-divertor, the molecules dominate the neutrals, such that $n_{n}^{gauge}=n_{mol}^{gauge}$; and further assuming that the gauge operates at the wall temperature ($T_{gauge}=T_{wall}$), and that $T_{mol}=T_{rot}$, we arrive at:
\begin{equation}
    n_{mol}^{div}=\frac{p_{n}}{k\sqrt{T_{wall}T_{rot}}}.
\end{equation}
It is assumed that the molecular density is constant throughout the neutral gas buffer.\par

The neutral gas buffer is assumed to exist only downstream the ionisation front. The poloidal distance between the target and the ionisation front is denoted as $L_{pol}$ and is obtained using the $D_2$ Fulcher emission front as a proxy for the ionisation front \cite{Wijkamp2023,Osborne2023,Verhaegh2023}. The volume of the neutral buffer is then $2\pi R L_{pol} w$, where $w$ is the width of the divertor leg and $R$ is the radius of the divertor leg. \\

Taking the various components together, combined with $\tau_{transport} = \frac{w}{v_{eff}}$, the power loss can be approximated as suggested in equation \ref{equ:power_loss}.

\begin{equation}\label{equ:power_loss}
P_{mol} \approx 5 \pi v_{eff} \sqrt{\frac{T_{rot}}{T_{wall}}} L_{pol} p_n R
\end{equation}

This provides an intuitive model for power losses from plasma-molecular collisions, which start after the detachment onset ($L_{pol} > 0$ m) and are driven by: 1) rotational temperature; 2) detachment front position; 3) divertor neutral pressure. \\

The model can be made more accurate by accounting for spatial variations of $T_{rot}$ in the divertor. Since $T_{rot}$ is experimentally obtained along a spectroscopic line of sight, chord-integrated power loss estimates can be obtained along each line of sight $P_{mol}^{LoS} [W/m^2]$, which can be integrated toroidally (e.g., $2 \pi R$ along the curve where the lines of sight intersect the separatrix, up to the ionisation front a distance $L_{pol}$ from the target) to obtain the total power loss, analogously to how ion sources and sinks are obtained experimentally \cite{Verhaegh2021}. \\

\begin{equation}
\begin{split}
P_{mol} &= 2 \pi\int_0^{L_{pol}}  R(s) P_{mol}^{LoS} (s) ds \\
P_{mol}^{LoS} (s) &\approx \frac{5p_n v_{eff} (s) \sqrt{T_{rot} (s)}}{2 \sqrt{T_{wall}}}
\end{split}
\end{equation}

The estimated total power losses from plasma-molecular collisions in both divertors, according to this estimate, are significant for the Super-X divertor and increase as the core density is increased due to: 1) a movement of the ionisation front further upstream; 2) increase in divertor neutral pressure; 3) increase in rotational temperature observed.\\

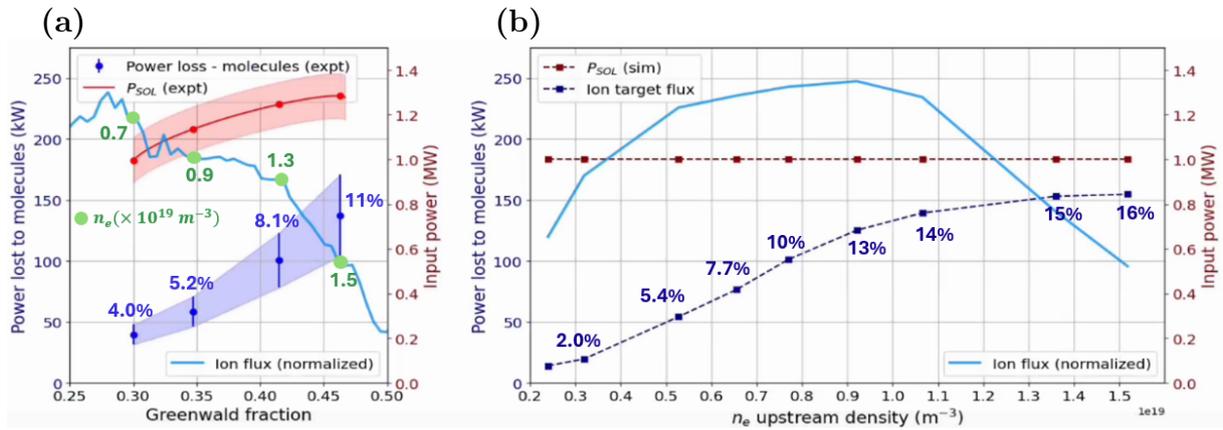
\begin{figure}[htbp]
    \centering
    \includegraphics[width=1\linewidth]{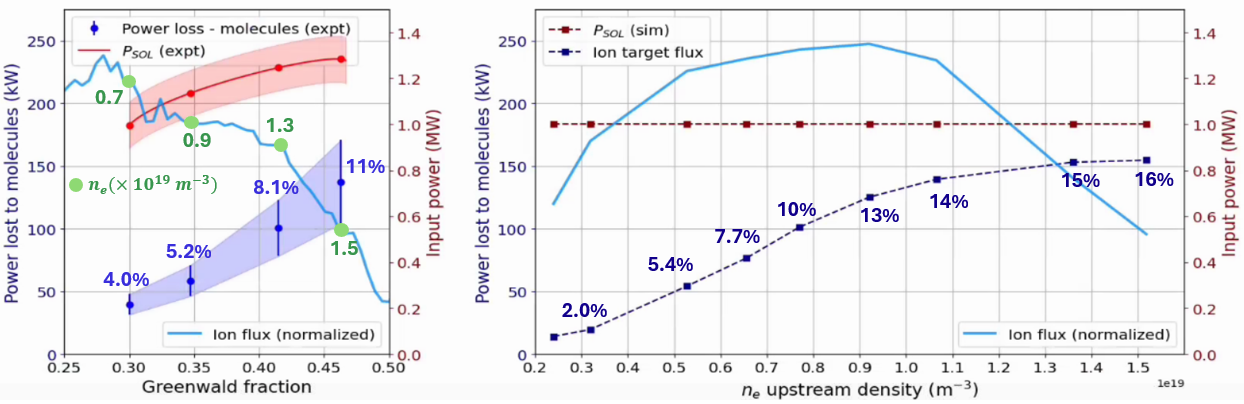}
    \begin{picture}(0,0)
    \put(-215, 165){\textbf{(a)}} 
    \put(-45, 165){\textbf{(b)}} 
    \end{picture}
    \caption{Plot showing the estimated power transferred to the molecules in the experiment (a), and the same value outputted from the simulations (b).  The percentages are the corresponding percentages of input power ($P_{SOL}$) lost in elastic collisions to the molecules.  For reference, the blue line shows the normalised target ion flux, and the green circles show the upstream electron density in the experiment.}
    \label{fig:Power_transfer}
\end{figure}

To apply this for MAST-U SXD discharge 48358 (see figure \ref{fig:Power_transfer}), we utilise our rotational temperature and neutral pressure divertor measurements.  The approximate variation of $L_{pol}$ with $f_{GW}$ in a MAST-U beam-heated density ramp is taken from \cite{KV-ne-Te-Lpol} (a repeat discharge of 48358) and we find a range $L_{pol}\approx 36-52\,\si{cm}$ corresponding to $f_{GW}\approx 0.30-0.46$.\\ 

To calculate $v_{eff}$:
\begin{equation}
    v_{eff}=\frac{\langle v_{th}\rangle}{\sqrt{2}wn\sigma}\,\,\,\text{where}\,\,\,\langle v_{th}\rangle\approx\sqrt{\frac{4k_bT_{rot}}{3m_{D_2}}}\,\,\,\text{(see Appendix)},
\end{equation}
we require several other parameters as well as the $T_{rot}$ measurements.  The cross-section is taken as $\sigma\approx6.5\times10^{-19}\,\si{m}^{2}$ from the AMJUEL data file 1.2.7 \cite{AMJUEL} (along with an appropriate scaling).  We take an approximate ion divertor density appropriate for calculating the mean free path of the molecules: $n_i\approx2\times10^{19}\,\si{m}^{-3}$ (see \cite{Nicola-physics}), $T_{wall}$ is taken as 300 K, and the width of the plasma leg (based on poloidal flux expansion and the upstream SOL width) is taken as $w\approx10\,\si{cm}$.\\

Using these measurements and parameters, in the most detached case, the estimated power losses are 100-170 kW, which is significant (8-13 \%) compared to an estimated input power ($P_{SOL}$) of 1.3 MW. This is in reasonable agreement with SOLPS-ITER simulations, which indicate similar power losses from elastic ion/molecule collisions (also shown on figure \ref{fig:Power_transfer}) and which suggest losses of up to $40-50\%$ of power remaining downstream of ionisation (not shown on plot). The power losses from plasma chemistry (e.g., MAD) are of similar order of magnitude ($\sim 15 \%$ $P_{SOL}$ \cite{Verhaegh2021b, Verhaegh2024}). Although this power ultimately reaches the vessel walls, it is spread over a much wider area than the target as the neutrals re-associate at the wall. Power losses from plasma-molecular interactions can thus be an important power loss channel in detached conditions. \\

\subsubsection{Momentum losses from elastic plasma-molecule collisions}

Volumetric momentum losses play a key role during detachment. Depending on interpretation \cite{Verhaegh2021b}, they are either required for detachment \cite{Stangeby2018}, or are essential for maintaining a high upstream pressure (or density) during detachment \cite{Krasheninnikov2017,Kukushkin2016,Pshenov2017}, required for reactor implementations.\\

Historically, such momentum losses were attributed to a dominance of atomic charge exchange over ionisation, which decays at low temperatures driving momentum losses \cite{Stangeby2018,Krasheninnikov2017,Self-Ewald}. However, more recent code investigations \cite{Moulton,Omkar} argue that the dominant momentum loss is driven by \emph{the same ion-molecule collisions} that result in an increase in rotational temperature and plasma power losses. This is supported by reduced modelling that argues the target temperature is a main driver of the divertor processes, since the molecular density is increased as target temperature is reduced \cite{Stangeby2017}. These theoretical findings provide a paradigm shift in understanding one of the most fundamental processes during detachment. There is no experimental validation of these processes yet.\\

However, if \emph{the same} plasma-molecule collisions result in 1) increased gas temperature; 2) plasma power losses ($\si{W/m}^3$); and 3) plasma-momentum losses ($\si{N/m}^3$), then knowledge about one of these processes could be used to obtain information about the other(s).\\

\begin{figure}[htbp]
    \centering
    \includegraphics[width=0.6\linewidth]{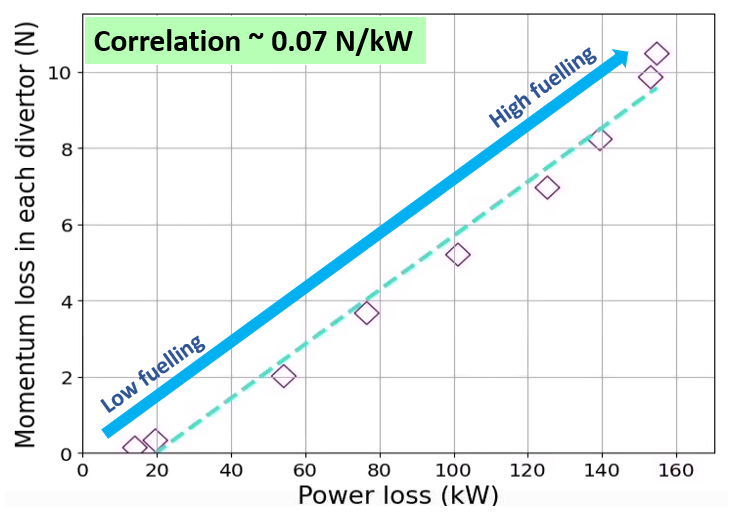}
    \caption{Figure indicating the correlation between power and momentum losses \textit{to the molecules} summed cell-by-cell in SOLPS-ITER simulations for MAST-Upgrade.}
    \label{fig:MomPower_calibration}
\end{figure}

To achieve this, we investigate the correlation between plasma power losses \textit{to the molecules} and volumetric plasma momentum losses \textit{to the molecules} from plasma-molecular collisions by summing the accumulated losses cell by cell from a range of SOLPS-ITER simulations.\\

A strong correlation between both quantities, shown in figure \ref{fig:MomPower_calibration}, is obtained:

\begin{equation}
    \frac{S^{LOS}_{mol}}{P^{LOS}_{mol}}\approx0.07\,\si{N/kW}.
\end{equation}

\begin{figure}[htbp]
    \centering
    \includegraphics[width=1\linewidth]{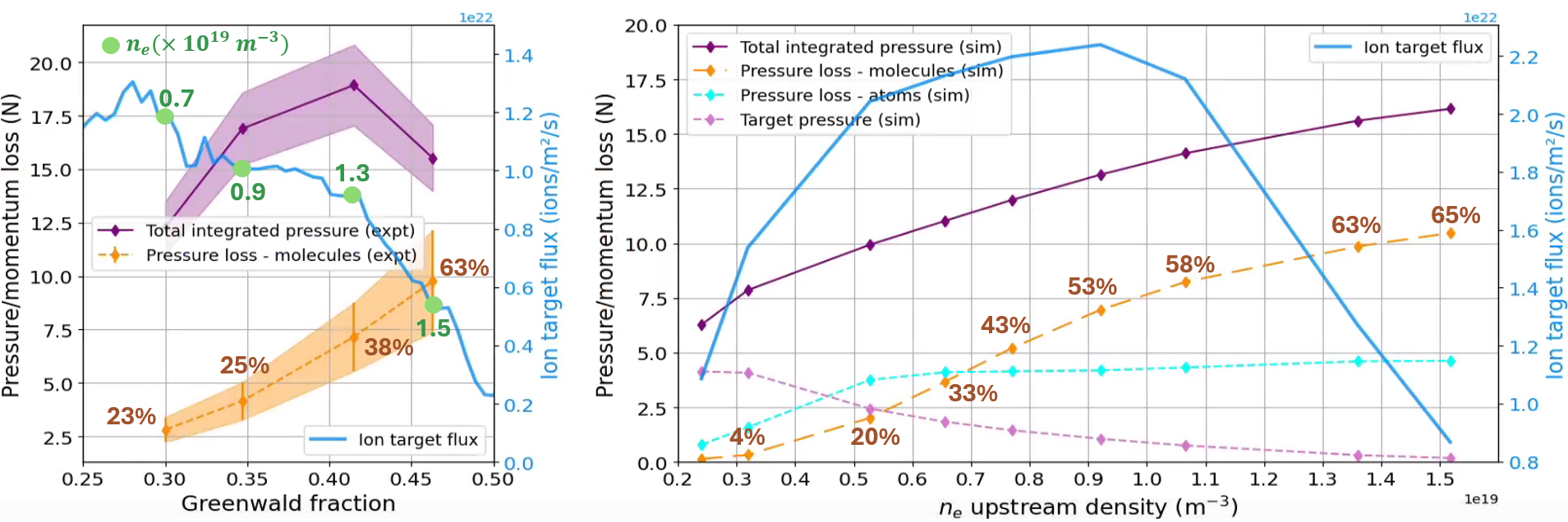}
    \begin{picture}(0,0)
        \put(-215, 165){\textbf{(a)}} 
        \put(-45, 165){\textbf{(b)}} 
    \end{picture}
    \caption{Inferred momentum losses in one divertor (using calibration from SOLP-ITER) compared against integrated pressure at mid-plane SOL for (a) experimental and (b) simulated SXD MAST-U data (including losses from atoms and target pressure. The percentages show the percentage of midplane pressure lost due to elastic collisions with the molecules.  The green circles in (a) indicate the upstream electron density.}
    \label{fig:MomLoss_Exp}
\end{figure}
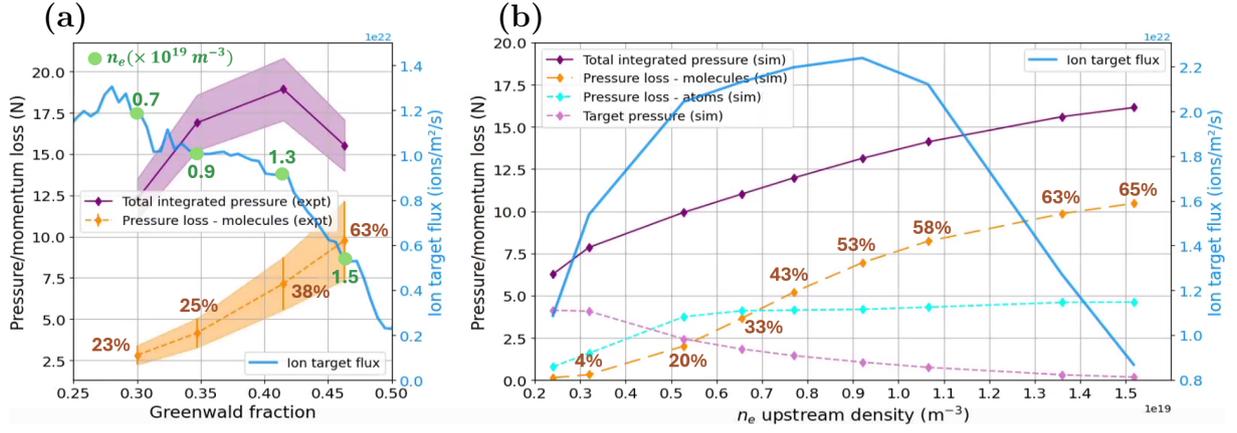

This implies that the chord-integrated power losses (modelled in section \ref{sec8}), combined with a ``calibration" inferred from the correlation between power and momentum losses in simulations, can be used to infer volumetric momentum losses experimentally, using the equation below. Applying these calibration coefficients, the divertor integrated momentum losses ($N$) from elastic ion-molecule collisions is estimated and shown in figure \ref{fig:MomLoss_Exp}.

\begin{equation}
S_{mol} = \frac{S_{mol}^{LoS}}{P_{mol}^{LoS}}\times2 \pi\int_0^{L_{pol}}  R(s) P_{mol}^{LoS} (s) ds
\end{equation}

Since the total momentum losses ($N$) are inferred (also summed for multiple flux tubes), they need to be brought into context by comparing them against the total integrated upstream pressure ($N$). The upstream profiles of $n_e$ and $T_e$ are integrated areally ($N/m^2 \rightarrow N$) to provide this comparison. The pressure upstream is approximated using an exponential profile, whose fall-off-length ($\lambda_p$) is related to the heat flux fall-off-length ($\lambda_q$) \cite{Faitsch,Stangeby2000}: $\lambda_p \approx \frac{7}{6} \lambda_q$ (see Appendix). Integrating this areally, to obtain an integrated pressure leads to: 
\begin{equation}
\begin{split}
p_{int} &= 2 \pi \int_{R_{sep}}^\infty r p(r) dr \\
p(r) &= 2 n_{e}^{sep} k T_{e}^{sep} \exp(-\frac{r-R_{sep}}{\lambda_p}) \\
\rightarrow p_{int} &\approx \frac{14}{3} \pi R_{sep} k T_{e}^{sep} n_{e}^{sep} \lambda_q.
\end{split}
\end{equation}

$\lambda_q$ is estimated from MAST scalings \cite{Harrison2013} and the separatrix density and temperature variation with $f_{GW}$ are obtained from previous analysis in \cite{KV-ne-Te-Lpol}.  With $R_{sep}\approx1.3\,\si{m}$ in MAST-U, we obtain (figure \ref{fig:MomLoss_Exp}a purple line):
\begin{equation}
    p_{int}\approx 12-20\,\si{N}.
\end{equation}
Comparing this areally integrated pressure to the volumetrically integrated momentum losses in each divertor (see figure \ref{fig:MomLoss_Exp}), we see that the majority of the momentum losses ($63\%$) in deeply detached conditions (high fuelling rate) could be explained by elastic ion-molecule collisions.  This is consistent with SOLPS-ITER predictions (also shown in figure \ref{fig:MomLoss_Exp}).  Interestingly, as also seen in figure \ref{fig:MomLoss_Exp}, SOLPS-ITER predicts that losses due to plasma-atom collisions saturate at relatively low upstream density (before detachment) while plasma-\textit{molecule} collisions become more and more dominant.\\

However, it should be noted that this experimental approach is approximate and is a first step. A more complete analysis of the upstream and target pressure - using both upstream and target measurements as well as 2D integrated Data Analysis \cite{Dan} - is left for future work, such that the inferred volumetric momentum losses can be compared against the total pressure loss obtained.

\subsubsection{Limitations of reduced model for power and momentum losses}

Although the experimental model provides estimates in quantitative agreement with SOLPS-ITER, there are a number of limitations. \\

First, the transport time of the molecules or effective transport velocity must be estimated, which cannot be accurately measured. However, it is not expected that this effective transport time changes significantly within the region downstream of the ionisation front and, therefore, it would not impact the trend of the power loss estimates obtained. Secondly, a constant molecular density is assumed in the neutral cloud based on neutral pressure measurements that are significantly away from the divertor plasma, which is not true.  The plasma density is also assumed constant whereas, ideally, a measurement is needed.  Thirdly, the neutral cloud is approximated as a box downstream the ionisation front. Although the neutral cloud is mostly concentrated downstream of the ionisation front, it also forms around the divertor plasma in these relatively low density/power (and thus large $\mathrm{D}_2$ mean-free-path) plasmas, as can be evidenced from $\mathrm{D}_2$ Fulcher emission occurring throughout the divertor leg, and not just at the target, in attached conditions. The above model may thus underestimate the impact of plasma-molecule collisions in attached conditions and is expected to become more accurate in more strongly detached conditions. Although the neutral buffer description could be made more accurate, this partially defeats the purpose of the reduced model: to provide an intuitive picture of the impact of plasma-molecular interactions, together with quantitative estimates. \\

An additional further limitation of the volumetric momentum loss estimates is that it relies on a calibration from a set of SOLPS-ITER simulations (with the same power input and geometry) to convert power losses into momentum losses. Further work and model expansion could remove the need for using such a calibration in the future by deriving this relationship between energy and momentum transfer using first principles.\\

The model would also benefit by being made more predictive.  The effect of poloidal flux expansion on power and momentum losses is clearly important, but the current interpretive model does not allow a direct comparison between $w$ and these losses (since there is an interdependency between $w$ and the rotational temperature achieved).  The model also currently has no upper bound for the total power or momentum loss.

\subsubsection{Implications of power and momentum loss models}

The reduced model used for power and momentum losses has wider implications. In this model, power and momentum losses from plasma-molecule collisions are driven by the ionisation front position, the divertor neutral pressure and gas temperature. Therefore, the detachment of the ionisation front from the target ($L_{pol} > 0$ m) results in power and momentum losses from plasma-molecule collisions in this model. Since these momentum losses are the dominant volumetric momentum losses, this implies that dominant volumetric momentum losses start to occur when the ionisation front detaches from the target.\\

In other work, it was argued that the detachment of the ionisation front from the target coincides with the onset of detachment \cite{Verhaegh2023, Verhaegh2023b} and other work used the ionisation front as a method to control detachment \cite{BobK}. This model provides a further physics basis for these arguments: real-time control of the ionisation front position, and density of the neutral buffer if possible, is predicted to enable real-time control of the plasma-molecule collisions that drive power and dominant momentum losses. \\

Recent work shows the \emph{additional volume available} drives the observed differences between longer-legged and shorter-legged divertors \cite{Verhaegh2024}, consistent with our model. MAST-U experiments indicated similar neutral pressures \cite{Verhaegh2024-arXiv-div-shaping} between the Super-X, Elongated and Conventional divertors. The rotational temperature at \emph{the same poloidal distance to the X-point} is similar between all three geometries. The difference between the three geometries therefore is primarily driven, according to this reduced model, by their difference in $L_{pol} \times R_t$: the longer-legged divertors feature a larger neutral buffer and therefore more plasma-molecular collisions resulting in power losses and volumetric momentum losses. Additionally, in that additional volume available, a colder divertor plasma is present leading to longer molecular lifetimes (i.e., higher rotational temperatures), leading to further enhancements in power and momentum losses.  \\

Assuming that plasma-molecular interactions are the dominant volumetric momentum loss channel, our reduced model can be used to predict differences in the momentum loss between different divertor shapes, which can be compared against the experimental results shown in \cite{Nicola-physics}. Our model predicts a $\sim 25$ \% reduction of the total momentum loss in the ED compared to the SXD (using $L_{pol}^{SXD}=0.38$ m, $L_{pol}^{ED} = 0.26$ m and $R_t^{SXD} = 1.35$ m, $R_t^{ED}=1.15$ m from \cite{Verhaegh2024-arXiv-div-shaping}, and accounting for a reduced $w$ by about 30\% in the ED). Assuming the volumetric momentum losses in the SXD ($f_{GW}=45 \%$) are e.g. $\sim 90 \%$ (if the effect of atoms is included as suggested by SOLPS-ITER in figure \ref{fig:MomLoss_Exp}), the momentum loss fraction ratio between the SXD and ED  $\frac{1-f_{mom, loss}^{SXD}}{1-f_{mom, loss}^{ED}}$ is expected to be $\sim 0.3$, which can be compared with the inferred momentum loss fraction ratios in \cite{Nicola-physics} ($0.15-0.5$). This showcases both the wider implications of the reduced model and provides some confidence in its validity. 

\subsection{Relevance and implications of this work}

Our results indicate that the rotational temperatures in tokamak divertors increase during detached conditions with a subsequent drop at the highest levels of detachment. The transition from attached to detached conditions seems to be the dominant driver of this increase for which we have developed an improved understanding. The rotational temperature is found to be correlated with electron density in many works \cite{Sergienko2013MolecularJET, Hollmann2006, Fujii2023Plasma-parameterDivertor} and our results do not contradict these findings. 
 We find that increasing rotational temperatures during detachment, in a wide range of divertor conditions, can be explained by longer molecular life times, resulting in more plasma-molecule collisions. Such collisions result in significant power and dominant volumetric momentum losses to the plasma.\\

Power balance is an often researched topic in power exhaust, as more than 90\% of $P_{SOL}$ must be dissipated in various reactor concepts \cite{DEMO-Zohm}. However, power balance studies often indicate that less than 90\% of $P_{SOL}$ can be accounted for \cite{EAST-90}. Due to the increased area over which the power loading from the neutral atoms to the wall is spread, the plasma power losses from plasma-molecule interactions - which can be tens of percent of $P_{SOL}$ - may not be well diagnosed in such power balance studies. This could explain part of the reason why obtaining power balance is challenging in fusion devices. \\

Reactors will operate at higher power and density conditions, where mean-free-paths of neutrals are reduced. However, a deeply detached long-legged divertor, with strong neutral baffling, will still have a significant neutral buffer \cite{BobK,Verhaegh2024}. Whether the importance of plasma-molecular interactions extends to reactor designs is unknown and requires further study. However, given the impact plasma-chemistry can have on reactor-scaled simulations \cite{Verhaegh2024}, it is likely that elastic plasma-molecule collisions play a key role in reactors with strongly/tightly baffled long-legged divertors.  This is because, while power losses resulting from plasma-molecule interaction, although significant, are relatively small compared to pre-detachment losses; and heat deposition from neutrals is unlikely to decrease (and may even increase) as a result of plasma-neutral interaction, we do see in this work that such collisions bring about large volumetric momentum losses.  Such momentum losses are critical in detachment to reduce ion target flux and prevent the collapse of upstream density.\\

This suspicion is consistent with SOLPS-ITER modelling of long-legged, tightly baffled divertors for reactors.  Investigating the momentum balance of characteristic STEP simulations \cite{Ryoko-STEP, Hudoba-STEP} suggests that the momentum loss driven by plasma-molecule collisions is either significant or dominant, depending on the level of detachment, consistent with the MAST-U results.\\

Meanwhile, it is important to note that fuel cycle requirements and detachment control may be impacted by plasma-neutral interactions.\\

Previous work shows disagreements between the impact of plasma-chemistry in simulations and experiments \cite{Verhaegh2023,Verhaegh2021b}, which are related to errors in the molecular charge exchange rates. The comparisons in this work suggest that there is a discrepancy between the vibrational distribution on which these rates are based, and that measured.  This may be due to inaccuracies in the collisional-radiative model used, or the vibrationally resolved setup, requiring further study. However, it is shown that the information that can be obtained on the vibrational distribution using $\mathrm{D}_2$ Fulcher emission is limited. Additional studies are likely required using active laser diagnostic techniques to measure the full vibrational distribution in the electronic ground state are required, which is being developed for linear devices such as Magnum-PSI \cite{Mag-PSI}. \\

Despite these disagreements with modelling, the observations of the impact of elastic plasma-molecule interactions are in agreement with modelling predictions. This provides some evidence that elastic plasma-molecule collisions, and their impact on the divertor plasma, are sufficiently accurately treated in exhaust simulation suites for TCV and MAST-U. However, further verification on such processes is required, particularly in higher power conditions and on metallic devices. Additionally, exhaust modelling suggests that the impact of plasma-molecule interactions during detachment is reduced in impurity seeded conditions \cite{Omkar,Smolders} - likely as the neutral pressure is reduced - which requires additional study. This information may be used in the future to further corroborate the models and physical picture proposed.

\section{Conclusion}\label{conc}

Investigating the rotational temperature in a range of different conditions on both TCV and MAST-U indicates, qualitatively, similar behaviour: the rotational temperature rises as the plasma becomes more deeply detached, until the rotational temperature becomes comparable to the plasma temperature. This observation holds true in multiple divertor shapes, external heating levels, baffled and unbaffled conditions, L- and H-mode, $\mathrm{H_2}$ and $\mathrm{D}_2$, as well as different fuelling locations. \\

A reduced model is derived to predict the observed increase of the rotational temperature (assuming it is approximately equal to the gas temperature). The gas temperature increases as the lifetime of the molecules is increased in a detached plasma, therefore the plasma undergoes more elastic ion-molecule collisions, resulting in higher gas temperatures. This finding is in agreement with SOLPS-ITER simulations: the molecular gas temperature is driven by ion-molecule collisions and increases during detachment, obtaining quantitatively similar values as the rotational temperatures observed. \\

Elastic collisions between the plasma and the molecules not only increase the molecular gas temperature, but also have significant impact on the divertor plasma according to SOLPS-ITER simulations. This work confirms the impact of elastic plasma-molecule collisions on the divertor state for the first time experimentally. \\

Using an approximate interpretive model, the plasma power losses from elastic molecule collisions are estimated using: 1) the rotational temperature (measure of increased collisions with the molecules and energy transfer per molecule); 2) the ionisation front position (size of the neutral buffer during detachment); 3) neutral pressure (estimate of the molecular density). Power losses from plasma-molecule collisions are significant in long-legged divertors, accounting for up to 10-15\% of $P_{SOL}$ in the deepest detached conditions. This provides further evidence of the benefit of long-legged, strongly baffled, divertors that are deeply detached and provides further insights into the significance of non-radiative power losses. \\

Combining this model with correlations between power and momentum losses obtained from SOLPS-ITER, it is shown that plasma-molecule collisions can result in dominant volumetric momentum losses. These momentum losses are expected to result in a strong target pressure reduction, essential for detachment whilst preventing an upstream pressure collapse. \\

With these results, this work shows novel insights into the importance of plasma-molecule collisions on detachment, which only become significant if there is a substantial amount of molecules (i.e. substantial neutral pressure) and a substantial spatial extent of the molecular cloud (i.e. detachment front significantly detached from the target). This provides further evidence and understanding of the benefit of long-legged, strongly baffled divertors and supports the limiting of target heat loading via real time ionisation front control.

\subsection*{Acknowledgements}
This work has received support from EPSRC Grants EP/W006839/1 and EP/S022430/1.  This work has been carried out within the framework of the EUROfusion Consortium, partially funded by the European Union via the Euratom Research and Training Programme (Grant Agreement No 101052200 — EUROfusion). The Swiss contribution to this work has been funded by the Swiss State Secretariat for Education, Research and Innovation (SERI). Views and opinions expressed are however those of the author(s) only and do not necessarily reflect those of the European Union, the European Commission or SERI. Neither the European Union nor the European Commission nor SERI can be held responsible for them.
\section{Appendix}\label{sec10}
\subsection{Derivation of reduced model for gas temperature increase}

We consider the collisions of ions with neutrals.  During such a collision, it can be demonstrated that the ion energy loss rate per unit volume is given by:

\begin{equation}\label{equ:Q}
    \frac{dE_{loss}/V}{dt}=8\left(\frac{2}{\pi}\right)^{\frac{1}{2}}n_i\,n_{mol}\,\sigma_{rep}\,k^{\frac{3}{2}}\frac{\{m_iM(m_iT+MT_i)\}^{\frac{1}{2}}}{(m_i+M)^2}(T_i-T),
\end{equation}
\\
where $m_i$, $M$, $T_i$, and $T$ are the ion mass, the target particle mass, the ion temperature and the target particle temperature respectively.  $k$ is the Boltzmann constant, and $\sigma_{rep}$ is a representative cross-section for the interaction \cite{Cravath}.
\\

Assuming that translational and rotational degrees of freedom are activated in the molecular cloud, then the energy of the molecular cloud per unit volume is given by:
\begin{equation}
    E_{cloud}/V = \frac{5}{2}n_{mol}\,T.
\end{equation}
Noting that the rate of energy gained by the molecular cloud by energy transfer from collisions will be equal to the rate of energy lost by the ions, then, in the case of $\mathrm{D^+}$ ions colliding with $\mathrm{D_2}$ molecules, (or for that matter $\mathrm{H^+}$ ions colliding with $\mathrm{H_2}$ molecules), equation \ref{equ:Q} becomes:
\begin{equation}\label{D2}
    \frac{d(\frac{5}{2}n_{mol}kT)}{dt}=\frac{16}{9\pi^{\frac{1}{2}}}n_i\,n_{mol}\,\sigma_{rep}\,k\,\left[\frac{k(T+2T_i)}{m_i}\right]^{\frac{1}{2}}(T_i-T).
\end{equation}
This leads to a collision energy transfer term:
\begin{equation}
    \frac{dT}{dt}\approx0.4\,n_i\,\sigma_{rep}\left[\frac{k(T+2T_i)}{m_i}\right]^{\frac{1}{2}}(T_i-T)
\end{equation}

We now consider loss terms due to destruction and transport.
\\

The rate of energy loss per unit volume due to destruction is given by:
\begin{equation}
    \frac{dE_{loss}/V}{dt}=-\frac{1}{\tau_{dest}}\frac{5}{2}n_{mol}kT+\frac{1}{\tau_{dest}}\frac{5}{2}n_{mol}kT_{wall},
\end{equation}
where $\frac{1}{\tau_{dest}}$ is the destruction frequency.  The second term replaces the lost molecules at $T_{wall}$ to ensure constant molecular density.
\\
\\
Similarly, the rate of energy loss per unit volume due to transport is given by:
\begin{equation}
    \frac{dE_{loss}/V}{dt}=-\frac{1}{\tau_{transp}}\frac{5}{2}n_{mol}kT+\frac{1}{\tau_{transp}}\frac{5}{2}n_{mol}kT_{wall},
\end{equation}
where $\tau_{transp}$ is the characteristic transit time of molecules in the plasma.
\\
\\
We define an effective loss time $\tau_{eff}$ encompassing both loss mechanisms:
\begin{equation}
    \frac{1}{\tau_{eff}}=\frac{1}{\tau_{dest}}+\frac{1}{\tau_{transp}},
\end{equation}
allowing us to arrive at the following differential equation:
\begin{equation}\label{final_model_de}
    \frac{dT}{dt}\approx0.4\,n_i\,\sigma_{rep}\left[\frac{k(T+2T_i)}{m_i}\right]^{\frac{1}{2}}(f\,T_i-T)-\frac{T}{\tau_{eff}}+\frac{T_{wall}}{\tau_{eff}},
\end{equation}
where $f$ is a scaler to account for inefficiency in the energy exchange term - in particular the requirement to maintain a molecular vibrational distribution.\\

This differential equation forms the basis for our reduced model.\\

To find a suitable representative cross-section in building our model (in keeping with the derivation of equation \ref{equ:Q} from \cite{Cravath}) we consider the energy exchange process at a given plasma temperature.  The energy moment associated with the collision operator has the kernel $\sigma v^{5}\,f_{ion}$, and since $v^5\,f_{ion}$ peaks sharply for a given ion energy, we take the cross-section corresponding to that peak as the representative cross-section.  We therefore utilise the cross-section data for ion-molecule collisions from the AMJUEL data files for EIRENE \cite{AMJUEL} (numerical fit H.1.2.7) rather than the $\langle\sigma v\rangle$ rate coefficient from the same data files (which would be inconsistent with this approach).\\

\subsection{Effective molecular travel speed}

The reduced model for the gas temperature increase requires taking into account the transit time of the molecules, $\tau_{transport}$, which depends on the distance the molecules must travel (e.g. the width of the divertor leg $w$) as well as the effective speed of the molecules travelling through the divertor leg, $v_{eff}$.\\

As the molecules travel through the divertor leg, their kinetic energy increases due to ion-molecule collisions. However, this does not necessarily dictate the effective speed of the molecules transiting through the divertor leg, as their direction is altered upon colliding with ions.  As such we assume the kinetics of the molecules are governed by a 1D random walk process and define:
\begin{equation}
    v_{eff}=\frac{w}{\tau_{transport}}.
\end{equation}
In this way, we introduce an approximate model to illustrate that the effective speed of the molecule travelling through the divertor is lower than the thermal velocity.\\

In this random walk, the effective distance travelled is:
\begin{equation}
    d_{eff}=w=\sqrt{N}\lambda,
\end{equation}
where $N$ is the number of steps taken (number of collisions) and $\lambda$ is the distance the molecule travels per step. We assume that this distance is equal to the mean-free-path, obtained between two collisions:
\begin{equation}
    \lambda = \frac{1}{\sqrt{2} n \sigma},
\end{equation}
where $n$ is the relevant particle density and $\sigma$ is the cross-section.\\

The actual distance travelled is: $D=N\lambda$ meaning that $D=\sqrt{N}w$.\\

If $d_{eff} > \lambda$, it will take:
\begin{equation}
    N=\left (\frac{d_{eff}}{\lambda}\right)^2=2w^2n^2\sigma^2
\end{equation}
 steps to travel $d_{eff}=w$.\\
 
 The time for one collision depends on the mean-free-path and the \emph{actual} travel speed of a molecule, $\langle v_{th}\rangle$, which we assume is determined by its kinetic energy:
 \begin{equation}
     t=\frac{\lambda}{\langle v_{th}\rangle},
 \end{equation}
 where:
\begin{equation}
    \langle v_{\text{th}} \rangle = \frac{1}{\Delta t} \int_{t_0}^{t_0 + \Delta t} \sqrt{\frac{3k_bT(t)}{m}} \, dt\approx\sqrt{\frac{3k_b\frac{4T_{final}}{9}}{m_{D_2}}}.
\end{equation}
We thus arrive at the following useful equations for the total transit time of the molecule:
\begin{equation}
    \tau_{transport} = N t = \frac{N \lambda}{\langle v_{th}\rangle} = \frac{\sqrt{2} w^2 n \sigma}{\langle v_{th}\rangle},
\end{equation}
and for the effective speed of the molecules:
\begin{equation}
    v_{eff}=\frac{\langle v_{th}\rangle}{\sqrt{2}wn\sigma}.
\end{equation}

It should be noted that this random walk approach is oversimplified (i.e., the path of the molecules is not 1D; if the root-mean-square distance travelled by the molecule exceeds the divertor leg, it is not true that it has transited through the leg). The point, however, is that the effective speed of the molecule travelling through the plasma is lower than the thermal velocity of the molecule due to the various collisions it undergoes with the plasma. The effective speeds and transit times obtained are consistent with other values reported in literature \cite{Yoneda2023SpectroscopicTokamaks}. \\

\subsection{Derivation of upstream pressure profile}

The upstream pressure profile can be estimated using a characteristic fall-off-width $\lambda_p$, based on upstream density and temperature profiles assuming an exponential decay. Furthermore, we assume $n_e=n_i$ and $T_e=T_i$ throughout. \\

\begin{equation}
    \begin{split}
        n_e (r) &= n_e^{sep} \exp(-\frac{r - R_{sep}}{\lambda_{n}}) \\
        T_e (r) &= T_e^{sep} \exp(-\frac{r-R_{sep}}{\lambda_{T}}) \\ \\
        \rightarrow p &= 2 p_e = 2 n_e^{sep} T_e^{sep} \exp(-\frac{r-R_{sep}}{\lambda_p}) \\
        \frac{1}{\lambda_p} &= \frac{1}{\lambda_n} + \frac{1}{\lambda_T}
    \end{split}
\end{equation}

Using the Two Point Model formulation, assuming that 1) all heat transport is driven by conductive transport; 2) no power losses occur upstream; 3) the upstream temperature is much larger than the target temperature ($T_e^{sep} \gg T_t$), we obtain $T_e^{sep} \propto q_\parallel^{2/7}$.  Assuming an exponential decay of both the temperature and heat flux upstream, using their characteristic fall-off-lengths, we obtain $\lambda_T = \frac{7}{2} \lambda_q$ \cite{Faitsch,Stangeby2000}.\\

Further assuming that the target heat flux, density and temperature profile have exactly the same \emph{shape} as the upstream profiles (i.e., $T_t (r) = T_t^{sep} \exp(-\frac{r - R_{sep}}{\lambda_T})$, $n_t (r) = n_t^{sep} \exp(-\frac{r - R_{sep}}{\lambda_n})$ and $q_{\parallel, t} (r) = q_{\parallel, t}^{sep} \exp(-\frac{r - R_{sep}}{\lambda_q})$), the Bohm sheath criterion ($q_{\parallel, t} \propto n_t T_t^{3/2} \propto p_t T_t^{1/2}$) can be used to relate $\lambda_{q}$, $\lambda_p$, $\lambda_T$. From this, we obtain: \\

\begin{equation}
    \frac{1}{\lambda_q} = \frac{1}{\lambda_p} + \frac{1}{2 \lambda_T} 
\end{equation}

Combining this with $\lambda_T = \frac{7}{2} \lambda_q$ results in $\lambda_p = \frac{7}{6} \lambda_q$.  \\

It is important to note that this method is approximate and contains various assumptions that are not valid (heat, pressure and temperature profiles change between upstream and the target, particularly in detached conditions). However, the point of this approach is to obtain rough estimates for the upstream pressure profile using parameters that can be obtained from literature. More quantitative future work, however, should feature a study of the upstream and target pressure profiles to relate the total pressure loss to the inferred volumetric momentum losses obtained from elastic ion-molecule collisions.

\clearpage

\printbibliography
\end{document}